\documentclass[runningheads]{llncs}
\usepackage{eccv}

\usepackage{eccvabbrv}

\usepackage[utf8]{inputenc} 
\usepackage[T1]{fontenc}    
\usepackage{hyperref}       
\usepackage{url}            
\usepackage{booktabs}       
\usepackage{amsfonts}       
\usepackage{nicefrac}           
\usepackage{lipsum}
\usepackage{graphicx}
\usepackage{doi}
\usepackage{hyperref}
\usepackage{url}
\usepackage{comment}
\usepackage{graphicx}
\usepackage{subcaption}
\usepackage{amssymb}
\usepackage{mathtools}
\usepackage{algorithm}
\usepackage[noend]{algpseudocode}
\usepackage{subcaption}
\usepackage{gnuplottex}
\usepackage{xparse}
\usepackage{float}

\usepackage{makecell}
\usepackage{array, multirow,graphicx}
\usepackage{float}
\usepackage{booktabs}
\usepackage{textcmds}
\usepackage[capitalize,noabbrev]{cleveref}
\usepackage{setspace}
\usepackage{makecell}
\usepackage{wrapfig}
\usepackage{authblk}
\usepackage{colortbl}
\definecolor{lightgray}{gray}{0.9}
\usepackage{booktabs}
\usepackage{multirow}
\usepackage{tabularx}

\usepackage[accsupp]{axessibility}

\usepackage{hyperref}

\usepackage{orcidlink}

\hypersetup{
pdftitle={2312.05496},
}

\begin{document}

\title{Implicit Steganography\\Beyond the Constraints of Modality} 

\titlerunning{Implicit Steganography Beyond the Constraints of Modality}

\author{Sojeong Song$^{\dagger,}$\inst{1}\orcidlink{0009-0005-9558-0072} \and
Seoyun Yang$^{\dagger,}$\inst{1}\orcidlink{0009-0004-7642-4034} \and \linebreak
Chang D. Yoo$^{\ddagger,}$\inst{1}\orcidlink{0000-0002-0756-7179} \and
Junmo Kim$^{\ddagger,}$\inst{1}\orcidlink{0000-0002-7174-7932}}

\authorrunning{S.~Song et al.}

\institute{${^{1}}$ Korea Advanced Institute of Science and Technology (KAIST), Daejeon,\\Republic of Korea \newline
  \email{akqjq4985@kaist.ac.kr}, \email{seoyun.yang@navercorp.com}
}

\maketitle

\renewcommand{\thefootnote}{\fnsymbol{footnote}}
\footnotetext[0]{$^{\dagger}$: Both authors have equally contributed.}
\footnotetext[0]{$^{\ddagger}$: Corresponding author.}
\renewcommand{\thefootnote}{\arabic{footnote}}

\begin{abstract}
Cross-modal steganography is committed to hiding secret information of one modality in another modality. Despite the advancement in the field of steganography by the introduction of deep learning, cross-modal steganography still remains to be a challenge to the field. The incompatibility between different modalities not only complicate the hiding process but also results in increased vulnerability to detection. To rectify these limitations, we present INRSteg, an innovative cross-modal steganography framework based on Implicit Neural Representations (INRs). We introduce a novel network allocating framework with a masked parameter update which facilitates hiding multiple data and enables cross modality across image, audio, video and 3D shape. Moreover, we eliminate the necessity of training a deep neural network and therefore substantially reduce the memory and computational cost and avoid domain adaptation issues. To the best of our knowledge, in the field of steganography, this is the first to introduce diverse modalities to both the secret and cover data. Detailed experiments in extreme modality settings demonstrate the flexibility, security, and robustness of INRSteg.
\end{abstract}

\section{Introduction}
\label{sec:intro}

Nowadays, an unprecedented volume of data is shared and stored across various digital platforms and the volume is bound to intensify even further in the future. This proliferation and accessibility of data contributes to convenience in our lives and enables innovation and advancement that weren't achievable in the past; however, it also brings a concerning challenge to the security of information. Steganography aims to hide the secret information inside a cover data resulting in a stego data, also called the container. Unlike cryptography, which investigates in hiding the data of interest in a coded form, steganography operates with imperceptible concealment so that the existence of the secret data is undetectable. For example, individuals can upload their stego data to the cloud and securely trade their private keys with specific parties, and companies can leverage steganography to exchange confidential information, such as new product designs or sensitive meeting recordings, while including ownership details in the data. From a government perspective, steganography can covertly transmit information to foreign spies by inserting confidential national security information into seemingly ordinary cover data.

Up to now, image is the most commonly used modality of cover data, as it is less sensitive to human perception and is comparatively simple to embed secret information. However, image steganography tends to have a limit on payload capacity, and is susceptible to detection. Deep learning methods \cite{NIPS2017_838e8afb, zhu2018hidden} have increased the security, but suffer from unstable extraction and high computational costs \cite{9335027}. Other modalities such as audio, video and 3D shape also undergo similar limitations \cite{LIU2019238, 9411680} and struggle with trade-offs between capacity and security. Moreover, for cross-modal steganography, where the modality of the secret and the cover data differs \cite{9015498, han2023deep, 9008510}, these limitations intensify due to the difficulties that emerge when the dimension or capacity of the secret data exceed that of the cover data or when the secret data contains temporal information.

Implicit neural representations (INRs) have recently attracted extensive attention as an alternative data representation, using continuous and differentiable functions to represent data via parameters of a neural network. The concept of INRs offers a more flexible and expressive approach than the conventional notion of discrete representations. Multiple modalities, as audio, image, video and 3d shape, are all presentable using a single neural network in various resolutions \cite{dupont2022data}. Moreover, INRs benefit in capacity by reducing the memory cost according to the network design \cite{lee2021metalearning, park2019deepsdf}.

\begin{figure}[ht]
\begin{center}
\centerline{\includegraphics[width=0.8\columnwidth]{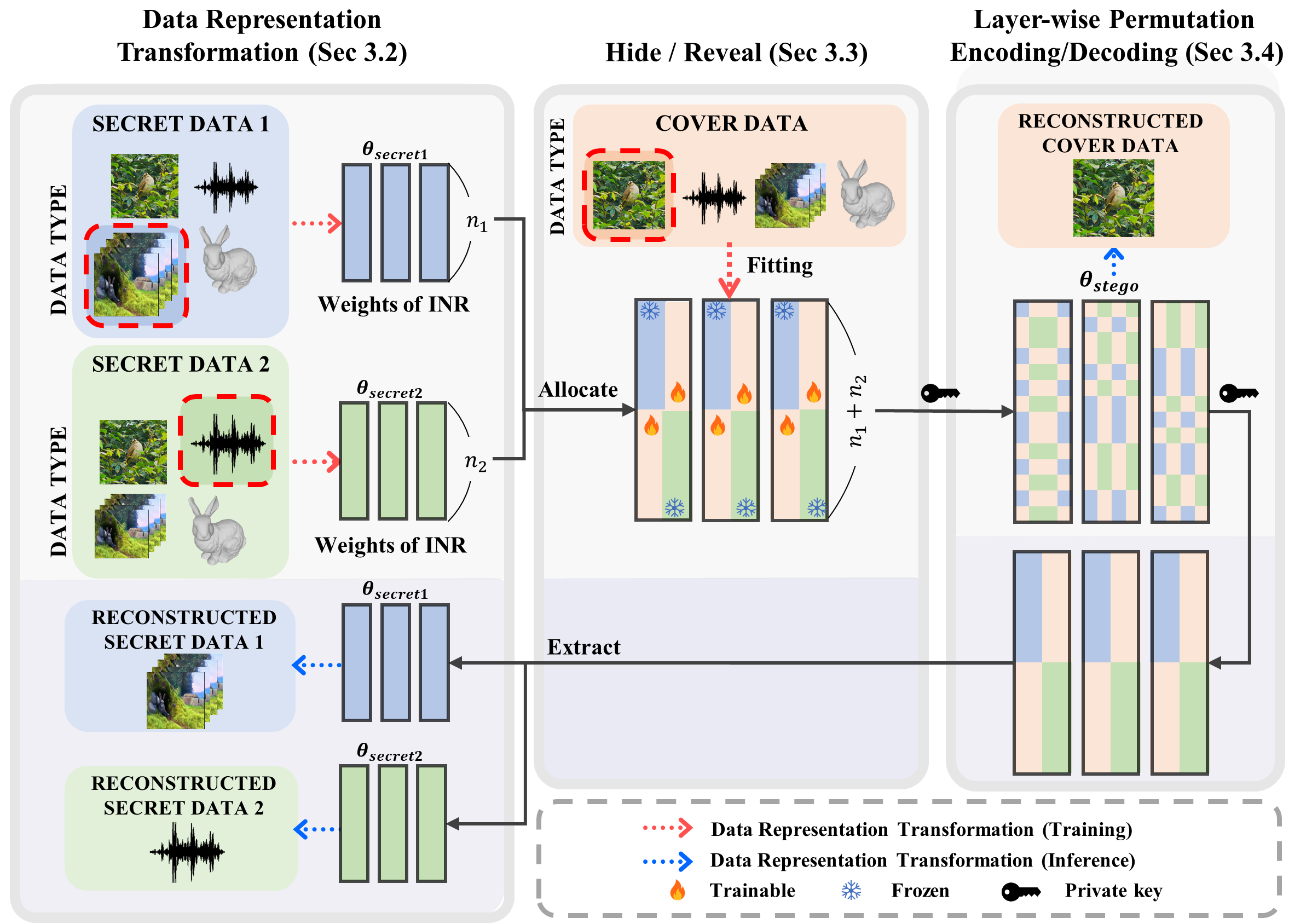}}
\caption{A general framework of INRSteg for hiding two types of secret data. All four data types can be transformed into INRs, and the red box shows which of the four data types is selected in this case. After the representation transformation of each secret data, the weights of the two INRs are allocated in a new MLP network. With the weight freezing technique, the weights from the secret data are fixed while the rest are fitted on the cover data. The existence of the secret data is concealed by permutation encoding via private key, which does not affect the reconstruction performance. The private key is then used to decode the permuted stego network, and the secret data are revealed by separating each MLP network.}
\label{fig: general framework}
\end{center}
\end{figure}

In this paper, we present INRSteg, the first cross-modal steganography framework available across a diverse range of modalities, including images, audio, video, and 3D shapes, for both the secret and cover data. By transforming the representation of all data to an implicit form, our framework takes advantage of the flexibility and high capacity inherent in INRs. The flexibility to all modalities enables cross-modal tasks that remained largely unexplored. Along with the versatility across the modalities, our framework is also capable of hiding multiple secret information into one cover data. In \cref{sec:experiment}, we conduct the hiding of both an audio and an video into a single image, a task which was previously considered highly challenging due to the limited capacity of images. Our framework therefore further enables expansion and intriguing adaptation of steganography into a wider range of fields. In the medical field, for example, private information as an endoscopy video and an audio recording of the heart sound can be hidden under an image of a patient.

To further enhance the security, layer-wise permutation of the INR network using a private key is additionally executed. Therefore, the secret INRs' network parameters can be well distributed within the stego INR's network. Revealing the secret data is executed by re-permuting and disassembling the weights of the stego INR. This procedure is not only simple but also guarantees a lossless revealing process as all neural representations of the secret data are fully recoverable. 

Additionally, our framework does not require any deep learning models to be trained for the hiding and revealing process, such as GAN \cite{shi2018ssgan, yang2018spatial} or diffusion models \cite{yu2024cross}. This excludes the need for any training data and therefore our framework can prevent any biases that may arise from using training datasets \cite{10.1145/3457607} and avoids domain shift issues, allowing datasets of all modalities to be applied without any additional pre-processing. Moreover, by eliminating the use of deep networks, our work substantially reduces both the computational and memory costs associated with training and deploying models which have become a significant global challenge.

The proposed framework, INRSteg, achieves flexible cross-modal steganography for multiple secret data by converting the data representation to neural networks. At the same time, we reduce computational and memory costs by avoiding the use of deep neural networks. For intra-modal steganography, we demonstrate that INRSteg outperforms prior image steganography models such as DeepMIH \cite{9676416} and DeepSteg \cite{NIPS2017_838e8afb}. We show INRSteg is robust in the real-world scenarios by applying quantization operation. These are our main contributions, and the details will be described in further sections: 

\renewcommand{\labelitemi}{$\bullet$}
\begin{itemize}
  \item {Introduce an Implicit Neural Representation (INR) steganography framework that flexibly adopts to a large range of modalities and further enables all cross-modal steganography.}
  \item {Demonstrate a neural network allocating procedure that enables cost-effective hiding and revealing of multiple secret data at once.}
  \item {Comprehensive investigation on distortion, security and capacity shows that INRSteg achieves state-of-the-art cross-modal steganography.}
  \item {Figure out that INRSteg is the most efficient compared to the previous steganography models and robust in real-world scenarios especially for quantization operation.}
\end{itemize}

\section{Related work}
\label{sec:related work}

\subsection{Steganography}
\label{Steganography}

Steganography is a technique that focuses on hiding confidential information by adeptly embedding data within other publicly accessible data, effectively concealing the very existence of the hidden message. When one is aware of this concealed data, it can be retrieved by a revealing process with an appropriate private key \cite{10.1007/3-540-39757-4_25}. A prominent approach in historical methods is the Least Significant Bit (LSB) based method, which takes advantage of the imperceptibility of small changes to the least significant bits of the data \cite{Jain2010ANI, 5209914}. With the rise of deep neural networks, the field of steganography has rapidly evolved to overcome the challenge of security and capacity. Deepsteg\cite{NIPS2017_838e8afb}, which is an early work of image steganography, attempts to use an encoder and decoder to determine where to place the secret information. CRoSS \cite{yu2024cross} utilizes diffusion models to only reveal the secret information when the accurate prompt is given. To enhance the capacity aspect while ensuring the security, invertible neural networks (INNs) have been actively adopted. In the image steganography field, HiNet\cite{Jing_2021_ICCV} incorporates INNs with wavelet domain to minimize distortion, and DeepMIH\cite{9676416} progresses to hide multiple images into one. LF-VSN \cite{mou2023large} further attempts to hide seven videos into one utilizing the INN.

Notwithstanding their innovations, a recurrent limitation across these previous methods is the trade-off between capacity and security. If the cover data is damaged in any way during the embedding process, it becomes susceptible to detection by steganalysis algorithms \cite{9153041, 7444146}. This vulnerability poses a significant limitation, especially when attempting to hide large volumes of data such as videos or 3d shapes. INRSteg cannot be detected by traditional steganlalysis tools and therefore ensures security while maintaining high capacity.

\subsection{Cross-modal steganography}
\label{Cross-modal steganography}
The evolution of steganography has also expanded to the concealment of information across different data modalities, such as embedding audio data within videos, which is called cross-modal steganography. \cite{9008510} embeds video information into audio signals by leveraging the human visual system's inability to recognize high-frequency flickering, allowing for the hidden video to be decoded by recording the sound from a speaker and processing it. \cite{9015498} utilizes a joint deep neural network architecture comprising two sub-models, one for embedding the audio into an image and another for decoding the image to retrieve the original audio. These methods achieve cross-modal steganography using deep learning techniques, but are still restricted on specified modalities.

Recent advancements in Implicit Neural Representations (INRs) have paved the way for a new paradigm in steganography\cite{li2022steganerf, han2023deep}. INRs have been leveraged to represent multiple modalities of data simultaneously. StegaNeRF \cite{li2022steganerf} hides data of various modalities at viewpoints when rendering NeRF. \cite{han2023deep} conceals audio, video, and 3D shape data within a cover image, leveraging RGB values as weights of the INRs. While these recent works represent a paradigm shift in steganography techniques, the capacity and security trade-off still remains an issue and the modality of the cover data is set to be singular. Our proposed method in this paper seeks to address and overcome this trade-off while maintaining the advantages of INRs and further eliminating the modality constraint on the cover data, demonstrating a significant advancement in the steganography research.

\subsection{Implicit neural representations}
\label{Implicit neural representations}
Implicit neural representations (INRs) is a rapidly emerging field where data is expressed in a continuous and differentiable manner through neural networks. This representation \cite{sitzmann2020implicit} is resolution-agnostic \cite{strumpler2022implicit} and is capable of compressing the data effectively \cite{dupont2021coin}. These advantages highlight various applications of INRs that show superior performance than that of the array-based representation. For example, neural radiance field (NeRF) \cite{mildenhall2021nerf} constructs a 5d scene representation of a 3d object through view-dependent synthesis. \cite{dupont2022data} proposes a general framework that feeds a modulated INR directly on the downstream models and shows competitive performance on multiple modalities. \cite{lee2021metalearning} employs meta-learning coupled with pruning techniques to efficiently scale implicit neural representations for large datasets.

\section{Method}
\label{sec:method}
\subsection{Framework}
\label{Framework}
Our research focuses on INR steganography, where the data representations are transformed into Implicit Neural Representations (INRs), such as SIREN \cite{sitzmann2020implicit}. \cref{fig: general framework} shows the general framework of our methodology, which is segmented into three sub-stages. In the \emph{Data Representation Transformation} stage, secret data are transformed into their corresponding Implicit Neural Representations, represented as ${\theta_{secret}}$. Subsequently, in the \emph{Hide/Reveal} stage, ${\theta_{stego}}$ is fitted to represent the cover data, ${x_{cover}}$, while preserving the parameters of ${\theta_{secret}}$. The \emph{Layer-wise Permutation Encoding/Decoding} stage introduces a layer-wise permutation mechanism, designed to avoid detection of the secret data's presence and location.

\subsection{Data Representation Transformation}
\label{Data Representation and Transformation}
In the first stage, we transform the secret data into Implicit Neural Representations (INRs) by fitting multi-layer perceptrons (MLP) with sine activation functions for each data. This network was first introduced by \cite{sitzmann2020implicit} as SIREN (Sinusoidal representations for neural networks) and is widely used in the INR research field. The structure of the MLP network has $n$ hidden layers of size ${D}$ and the output is represented as follows:
\begin{equation}
    \label{SIREN MLP}
    \begin{split}
        &\mathbf{y}=\mathbf{W}^{(n)}\left(g_{n-1} \circ\cdot\cdot\cdot\circ g_1 \circ g_0\right) (\mathbf{x}_0)+\mathbf{b}^{(n)} {,}\quad\quad\quad\quad\quad\quad\quad\quad\\
        &\quad\quad\quad\quad\quad\quad\quad\quad\operatorname{where}\:\mathbf{x}_{i+1}= g_i\left(\mathbf{x}_i\right)=\sigma \left(\mathbf{W}^{(i)} \mathbf{x}_i+\mathbf{b}^{(i)}\right),
    \end{split}
    \raisetag{2.5\normalbaselineskip}
\end{equation}
where ${\circ}$ is a function composition of fully-connected layers. In the equation, $\mathbf{x_i}$ is the input of the $i^{\text {th}}$ layer for ${i}\in\{0, 1, 2, ..., n\}$, $\mathbf{y}\in\mathbb{R}^{O}$ is the output value corresponding to $\mathbf{x_0}\in\mathbb{R}^{I}$, and $g_i\colon\mathbb{R}^{D}\rightarrow\mathbb{R}^{D}$. For our experiments, input and output dimensions ${(I, O)}$ are ${(1, 1)}$, ${(2, 3)}$, ${(3, 1)}$, ${(3, 3)}$ for audio, image, 3D shapes, and video, respectively. If the input or output dimension of the secret INRs exceeds that of the cover INR, our method employs a strategic reduction in the number of hidden layers within the secret INR. Additionally, as the dimension of the hidden layer $D$ is fixed for each cover modality, the secret INRs' hidden layer dimensions are set accordingly.
    
In general, given a set $S$ of inputs $\textbf{x}$$\in$$\mathbb{R}^{I}$ and the corresponding outputs $\textbf{y}$$\in$$\mathbb{R}^{O}$, the loss function is given by 
\begin{equation}
\label{e1}
{\mathcal{L}_{recon}}{\left(\theta\right)}{=}{\sum_{(\textbf{x},\textbf{y}) \in \mathcal{S}}\left\|f_{\theta}\left(\textbf{x}\right)-\textbf{y}\right\|_2^2}
\end{equation}, 
where $f_{\theta}$ is the neural representation of the data. For example, given a 2D image, the set $\mathcal{S}$ consists of 2D coordinates $\textbf{x}$$\in$$\mathbb{R}^{2}$ and the corresponding RGB values $\textbf{y}$$\in$$\mathbb{R}^{3}$. INRs enable a variety of data types to be expressed in this unified network architecture. This benefits the overall security paradigm of our steganography framework by hiding the underlying data types.

\subsection{Hide and Reveal}
\label{Hiding Stage}

\begin{wrapfigure}{R}{0.5\textwidth}
\centering
\includegraphics[width=0.5\textwidth]{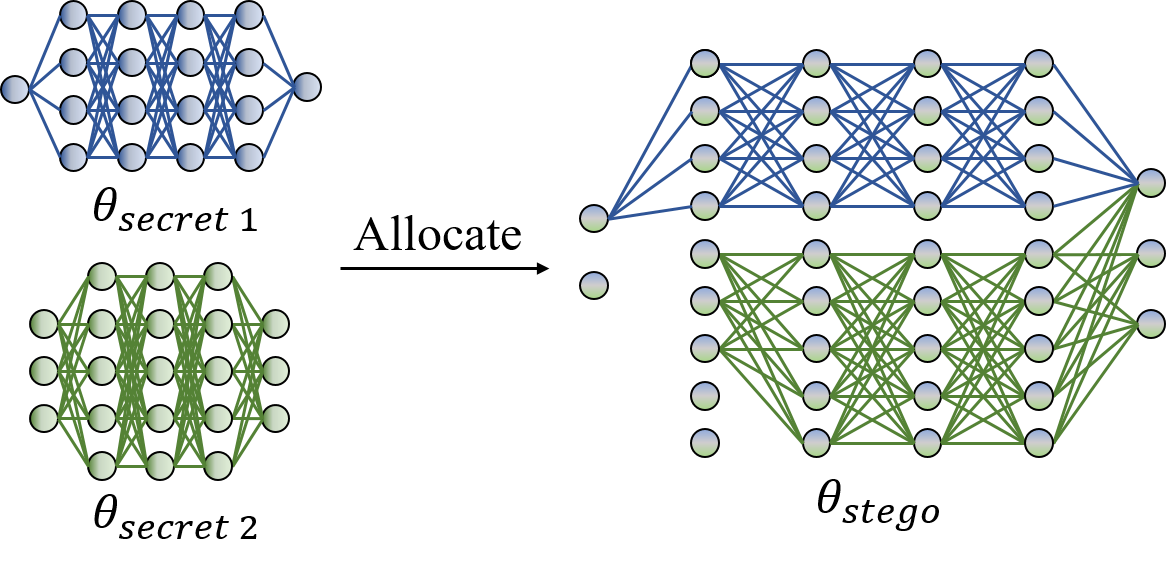}
\caption{An example of allocating two secret INRs, ${\theta_{secret1}}$ and ${\theta_{secret2}}$, into the stego INR, ${\theta_{stego}}$. For ${\theta_{secret1}}$, ${\theta_{secret2}}$, and ${\theta_{stego}}$, ${(I_1=1,}$ ${O_1=1,}$ ${n_1=4,}$ ${D_1=4)}$, ${(I_2=3,}$ ${O_2=3,}$ ${n_2=3,}$ ${D_2=5)}$, and ${(I=2,}$ ${O=3,}$ ${n=4,}$ ${D = 9)}$, respectively.}
\label{fig: inrmix}
\end{wrapfigure}

In this section, we explain our main idea associated with the hiding process. The secret INRs, ${\theta_{secret}}$, undergo a process of allocation and fitting to the cover data, ${x_{cover}}$. Within the weight space, only the diagonal blocks of weight matrices are occupied by the parameters of ${\theta_{secret}}$, so that the non-diagonal parameters can be fitted to ${x_{cover}}$, while ensuring the diagonal block's weight values to remain unaltered.

First, we explain when the number of secret data, $N$, is two without losing generality for hiding multiple data. The number of input nodes, output nodes, hidden layers, and the dimension of the hidden layers of ${\theta_{secret1}}$ and ${\theta_{secret2}}$ are each set to ${(I_1, O_1, n_1, D_1)}$, ${(I_{2}, O_{2}, n_2, D_2)}$, respectively. And ${(I, O, n, D)}$ for the stego INR network, ${\theta_{stego}}$, where $D_1 + D_2$ is set to equal $D$. Now, we substitute the parameters of the initialized ${\theta_{stego}}$ by concatenating ${\theta_{secret1}}$ and ${\theta_{secret2}}$ as shown in \cref{fig: inrmix}. Note that the initialized parameters of ${\theta_{stego}}$ are omitted for clarity. After the substitution, the parameter of ${\theta_{stego}}$ is as follows:
\begin{equation}
\footnotesize
\renewcommand{\arraystretch}{1.3}
\setlength{\arraycolsep}{5pt}
\mathbf{W}^{(i)}=\left(\begin{array}{ll}
\mathbf{W}_1^{(i)} & \mathbf{W}_{01}^{(i)} \\
\mathbf{W}_{02}^{(i)} & \mathbf{W}_2^{(i)}
\end{array}\right), \mathbf{b}^{(i)}=\left(\begin{array}{l}
\mathbf{b}_1^{(i)} \\
\mathbf{b}_2^{(i)}
\end{array}\right).
\raisetag{4.5\normalbaselineskip}
\label{INRconcate}
\end{equation}
Here, $\mathbf{W}^{(i)}$, $\mathbf{W}_1^{(i)}$, $\mathbf{W}_2^{(i)}$ and $\mathbf{b}^{(i)}$, $\mathbf{b}_1^{(i)}$, $\mathbf{b}_2^{(i)}$ are the $i^{\text {th}}$ layer's weight matrices and bias vectors of ${\theta_{stego}}$, ${\theta_{secret1}}$ and ${\theta_{secret2}}$, where ${i}\in\{0, ..., n\}$. If ${\theta_{secret1}}$ or ${\theta_{secret2}}$ have fewer hidden layers compared to ${\theta_{stego}}$, the parameters of ${\theta_{stego}}$ are not substituted. $\mathbf{W}_{01}^{(i)}$ and $\mathbf{W}_{02}^{(i)}$ are the remaining parameters of ${\theta_{stego}}$.

Now, ${\theta_{stego}}$ is fitted to ${x_{cover}}$ through a selective parameter update using a binary mask, so that the parameters of ${\theta_{secret1}}$ and ${\theta_{secret2}}$ are preserved while the remaining weights get updated. After successfully updating the weight space, the reconstructed discrete representation depicts ${x_{cover}}$ and all secret data can be retrieved from ${\theta_{stego}}$.

For ${N > 2}$, the hiding process is expanded in the same manner, so we further explain when ${N = 1}$. When there is only one secret data, the parameters of ${\theta_{secret1}}$ can be freely allocated in ${\theta_{stego}}$. Suppose that ${\theta_{secret1}}$ is allocated in the middle of ${\theta_{stego}}$, then the weight space of ${\theta_{stego}}$ is as follows:
\begin{equation}
\footnotesize
\renewcommand{\arraystretch}{1.3}
\setlength{\arraycolsep}{5pt}
\mathbf{W}^{(i)}=\left(\begin{array}{lll}
\mathbf{W}_{01}^{(i)} & \mathbf{W}_{02}^{(i)} & \mathbf{W}_{03}^{(i)} \\
\mathbf{W}_{04}^{(i)} & \mathbf{W}_1^{(i)} & \mathbf{W}_{05}^{(i)} \\
\mathbf{W}_{06}^{(i)} & \mathbf{W}_{07}^{(i)} & \mathbf{W}_{08}^{(i)}
\end{array}\right), \mathbf{b}^{(i)}=\left(\begin{array}{c}
\mathbf{b}_{01}^{(i)} \\
\mathbf{b}_1^{(i)} \\
\mathbf{b}_{02}^{(i)}
\end{array}\right),
\raisetag{4.5\normalbaselineskip}
\label{INRconcate_2}
\end{equation}
where $\mathbf{W}^{(i)}$, $\mathbf{W}_1^{(i)}$ and $\mathbf{b}^{(i)}$, $\mathbf{b}_1^{(i)}$ are the $i^{\text {th}}$ layer's weight matrices and bias vectors of ${\theta_{stego}}$ and ${\theta_{secret1}}$, for ${i}\in\{0, ..., n\}$. $\mathbf{W}_{01}^{(i)}, \mathbf{W}_{02}^{(i)}, \cdots, \mathbf{W}_{08}^{(i)}, \mathbf{b}_{01}^{(i)}$, and $\mathbf{b}_{02}^{(i)}$ are the remaining parameters of ${\theta_{stego}}$. This network then goes through the same processes as when there are multiple secret data. 

The inherent flexibility of our approach is highlighted by its capability to insert diverse data types, in varying amounts, at arbitrary desired positions. This versatility addresses and mitigates capacity constraints.

\subsection{Layer-wise Permutation Encoding and Decoding}
\label{Permutation Encoding}
To enhance security of our method, we strategically permute the nodes within each layer of the stego network. This layer-wise permutation allows the parameters of the secret INRs to be distributed throughout $\theta_{stego}$, making the parameters indistinguishable and thus undetectable. In this section, we demonstrate how the permutation is executed via one 128-bit private key ${\rho}$, utilizing a cryptographic key derivation function (KDF), which derives one or more secret keys from a master key. First, the private key is randomly selected and is employed to generate ${n}$ secret keys for the ${n}$ hidden layers of $\theta_{stego}$. Subsequently, the nodes of each layer are permuted according to the secret key assigned to each layer. The total number of possible permutation variations per INR is $(\mathbf{D}!)^{n}$, where ${D}$ is the dimension of the hidden layer. This shows that the security implications of this permutation intensify, as the possible combinations of permutation grow exponentially with the number of hidden layers in the $\theta_{stego}$.

It is crucial to emphasize that the overall functionality of the neural network remains unchanged throughout this operation. This operation leverages the inherent property of permutation invariance exhibited by Multi-Layer Perceptrons (MLPs) within a single layer. As INRs are also MLP networks, all permuted networks are identical to the original network. Therefore, $\theta_{stego}$ is invariant to the layer-wise permutation, whereas extracting the secret INR networks becomes impossible without the private key. 

\section{Experiment}
\label{sec:experiment}
We explore the performance of INRSteg focusing on its resilience to distortion, capacity for information hiding, and security against detection. Distortion measures both the indistinguishability between the cover data and the reconstructed data from stego INR, and the performance drop during the hiding and revealing stage of the secret data. Capacity  refers to the amount of hidden information that can be privately embedded within the cover data. Security indicates the extent to which the secret data can avoid detection when subjected to steganalysis.

To thoroughly evaluate the distortion, we employ several metrics tailored to different modalities, which are organized in Appendix A, as well as visualization. In terms of capacity, we present various cross-modal steganography experiments and compare the model size with other previous steganography models. For security, we compute the detection accuracy using image steganalysis models, SiaStegNet \cite{9153041} and XuNet \cite{xu2017deep}, and visualize weight distribution to show the effects of permutation operation. 

In our groundbreaking research, we highlight three pivotal aspects of our experimental evaluation of INRSteg, a novel steganography method that sets new benchmarks in data hiding techniques:

\renewcommand{\labelitemi}{$\bullet$}
\begin{itemize}
  \item {\textbf{Cross-Modal Steganography} We initiated pioneering experiments in cross-modal steganography, demonstrating our method's unique capability to privately embed audio and video data within images. Comprehensive ablation study demonstrates that INRSteg is generally applicable across all data modalities achieving high-quality concealment(\cref{experiment: Cross-modal Steganography}).}
  \item {\textbf{Intra-Modal Steganography} We conduct experiments of hiding data within the same modality, specifically embedding images within images. This approach allowed us to conduct a comparative analysis of distortion and security metrics against those reported in existing literature, showcasing the advanced capabilities of INRSteg. For ablation study, we conduct multi-image steganography experiments varying the amount of secret data(\cref{experiment: Intra-modal Steganography}).}
  \item {\textbf{Efficiency and Robustness Analysis} To show the efficiency of INRSteg, we perform computational cost comparisons with five existing steganography models. We test the robustness of INRSteg by simulating a challenging scenario where network weights undergo quantization. It demonstrates that our method maintains high performance and robustness even under harsh conditions, which ensures secure and reliable steganography across a wide range of applications(\cref{experiment: Efficiency and Robustness Analysis}).}
\end{itemize}

\begin{table}[h]
  \caption{Results of Cross-modal steganography experiment. VoxCeleb2 and ImageNet dataset are used for secret data and cover data, respectively. ``Secret/Revealed'' indicates the recovery performance of secret data. ``Cover/Stego'' indicates the reconstruction performance from stego INR.}
  \label{tab:cross-modal voxceleb2}
  \centering
  \scriptsize
  \renewcommand{\arraystretch}{1.05}
  \begin{tabularx}{\textwidth}{@{\extracolsep{\fill}}c*{4}{>{\centering\arraybackslash}X}|*{2}{>{\centering\arraybackslash}X}@{}}
    \toprule
    & \multicolumn{4}{c|}{{Secret/Revealed}} & \multicolumn{2}{c}{{Cover/Stego}} \\
    \cmidrule(r){2-5} \cmidrule(r){6-7}
    & \multicolumn{4}{c|}{{VoxCeleb2}} & \multicolumn{2}{c}{{ImageNet}} \\
    \cmidrule(r){2-5} \cmidrule(r){6-7}
    & \multicolumn{2}{c}{{Audio}} & \multicolumn{2}{c|}{{Video}} & \multicolumn{2}{c}{{Image}} \\
    \cmidrule(r){2-3} \cmidrule(r){4-5} \cmidrule(r){6-7}
    & SNR $\uparrow$ & MAE $\downarrow$ & PSNR $\uparrow$ & APD $\downarrow$ & PSNR $\uparrow$ & RMSE $\downarrow$ \\
    & 37.558(${\pm}$5.4) & 0.314(${\pm}$0.3) & 41.448(${\pm}$3.1) & 1.349(${\pm}$0.5) & 39.258(${\pm}$4.4) & 2.948(${\pm}$1.5) \\
    \bottomrule
  \end{tabularx}
\end{table}

\subsection{Cross-modal Steganography}
\label{experiment: Cross-modal Steganography}

\subsubsection{ImageNet and VoxCeleb2}
\begin{figure}[tb]
  \centering
  \includegraphics[width=0.6\columnwidth]{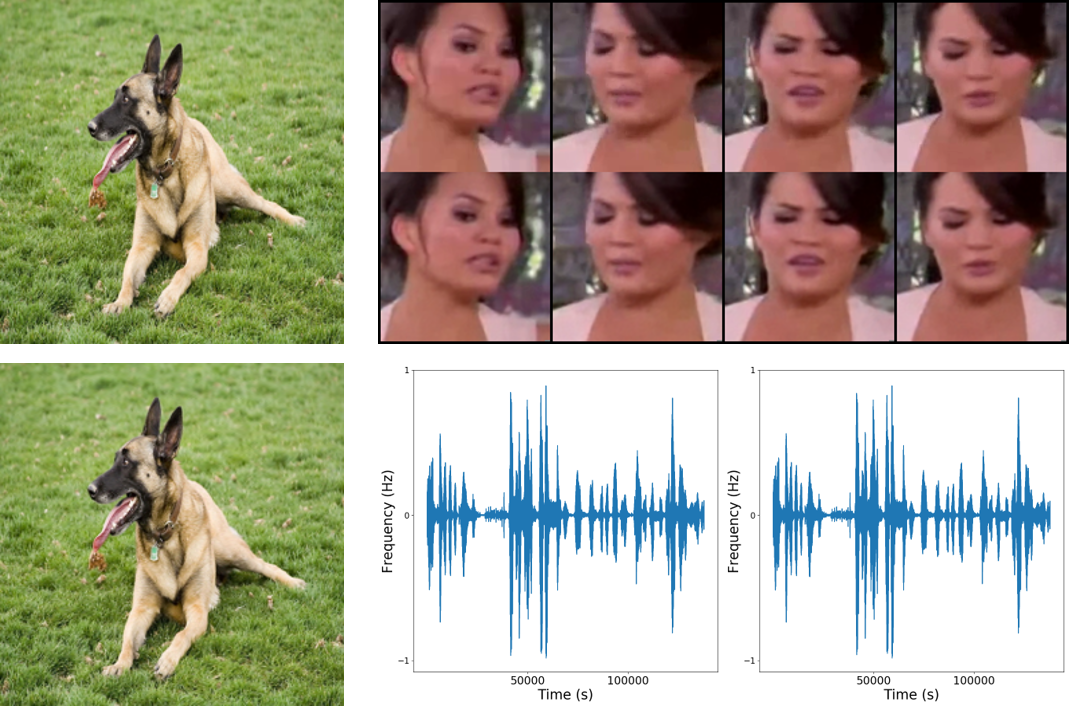}
  \caption{Visualization results of cross-modal steganography. Can you identify which image is the cover and the reconstructed cover data? For the video and audio examples, can you determine which are the secret and revealed secret data? For images, upper one is cover data and below is reconstructed cover data. For video, upper row is secret data and below is revealed secret data. For audio, the left is secret data and the right is revealed secret data.}
  \label{fig: cross_modal_vis}
\end{figure}

For cross-modal steganography experiment, we conduct a pioneering steganography experiment that embeds the VoxCeleb2 dataset within images from ImageNet dataset. VoxCeleb2 dataset has both video and audio files, thereby enabling simultaneous playback of visual and auditory data. Using our INRSteg method, we leverage images from ImageNet dataset as cover data to hide video and audio data at the same time. \cref{tab:cross-modal voxceleb2} shows the distortion performance for secret/revealed and cover/stego performance demonstrating the efficacy of our approach in preserving the content of cover and the embedded data. For the secret and the revealed secret data pair, as our framework goes through lossless recovery in terms of INR, the secret/revealed secret performance is solely dependent on the data representation transformation phase. Moreover, \cref{fig: cross_modal_vis} shows visualization of the reconstructed data, recovery of video and audio secret data. There is no noticeable quality degradation, which implies successful results of cross-modal steganography. Notably, cross-modal steganography is unprecedented, representing a significant improvement in the field of steganography. More reconstruction results using other modalities can be seen in Appendix C, showing successful multi-data steganography with almost no visible difference for both 3D shape and video. Additionally, there are visualization results of weight space after steganography and comparison results of weight distribution after permutation operation in Appendix D, demonstrating the security of INRSteg.

\subsubsection{Ablation study}
For ablation study, we try to demonstrate that INRSteg is generally applicable across all data modalities for steganography. Therefore, we design experiments that cover all combinations of data modalities where cover data type is image, audio, or video and secret data type is image, audio, 3D shape, or video. The distortion performance of all single-data cross-modal steganography tasks can be seen in \cref{tab:cross-modal single-data}, demonstrating achievement of exceptional performance for all cases. For additional reconstruction results, we show them in Appendix C. Recovery performance results for secret data are in Appendix E. A deeper analysis of the results presented in \cref{tab:cross-modal single-data} reveals two key factors that influence improved cover/stego performance: the modality of the secret and cover data, and the capacity of the neural network utilized for the stego INR. We observe that when the modalities of the secret and cover data match, the stego INR demonstrates enhanced accuracy in fitting to the cover data. This phenomenon is attributed to the intrinsic properties of INRs, where the weight spaces of neural networks exhibit similarities when representing with same modality data\cite{tancik2021learned}. Moreover, leveraging the capabilities of INRs, employing a larger neural network for the cover data fitting process enhances the quality of cover data reconstruction. This can be shown in \cref{tab:cross-modal single-data}, where the performance of hiding 3D shapes in images outperforms hiding images in images. This property allows flexible control of the network size for secret and cover data based on the required recovery quality and resource space constraints in the real world. An additional ablation study about the effectiveness of the padding ratio is proposed in Appendix B.

\begin{table}[h!]
  \centering
  \caption{Cover/Stego performance of single-data steganography. The best results are \textbf{highlighted} for each cover modality.}
  \label{tab:cross-modal single-data}
  \scriptsize
  \renewcommand{\arraystretch}{1.1}
  \setlength{\tabcolsep}{4pt}
  \begin{tabular}{@{}c|cc|cc|cc@{}}
    \toprule
    \multicolumn{1}{c|}{} & \multicolumn{6}{c}{{Cover type}} \\
    \cmidrule(l){2-7} 
    \multicolumn{1}{c|}{} & \multicolumn{2}{c|}{{Image}} & \multicolumn{2}{c|}{{Audio}} & \multicolumn{2}{c}{{Video}} \\
    \cmidrule(l){2-7} 
    {Secret type} & { PSNR $\uparrow$ } & { RMSE $\downarrow$ } & { SNR $\uparrow$ } & { MAE $\downarrow$ } & { PSNR $\uparrow$ } & { APD $\downarrow$ } \\
    \midrule
    {Image} & 62.34 & 0.204 & 32.63 & 0.599 & 35.26 & 3.315 \\
    {Audio} & 63.16 & 0.179 & \textbf{41.42} & \textbf{0.268} & 34.69 & 3.574 \\
    {Video} & 56.18 & 0.408 & 27.20 & 0.931 & \textbf{41.26} & \textbf{1.532} \\
    {3D shape} & \textbf{92.95} & \textbf{0.000} & 23.12 & 1.403 & 38.54 & 2.244 \\
    \bottomrule
  \end{tabular}
\end{table}

\subsection{Intra-modal Steganography}
\label{experiment: Intra-modal Steganography}
\subsubsection{Image steganography}
In this section, we compare INRSteg to existing image-to-image steganography methods to show that our framework also excels in intra-modal tasks. \cref{table: image-to-image steganography comparison} compares the performance of INRSteg with other deep learning based image-to-image steganography methods, DeepMIH \cite{9676416} and an improved DeepSteg \cite{NIPS2017_838e8afb}. INRSteg outperforms on all metrics showing that our framework achieves state-of-the-art performance. For hiding two secret images into one image, the results in Appendix F, and \cref{fig: image difference map} further visualizes the cover/stego and secret/revealed secret image pairs for two images into one steganography. We continue to compare our results with DeepMIH using difference maps each enhanced by 10, 20, and 30 times. For INRSteg, nearly no visual differences exist between the original and the revealed images even when the difference map is enhanced 30 times. For DeepMIH, on the other hand, we can notice obvious differences for all difference maps when comparing cover/stego and secret1,2/revealed secret 1,2 pairs. 
\begin{figure}[h]
  \centering
\includegraphics[width=0.9\columnwidth]{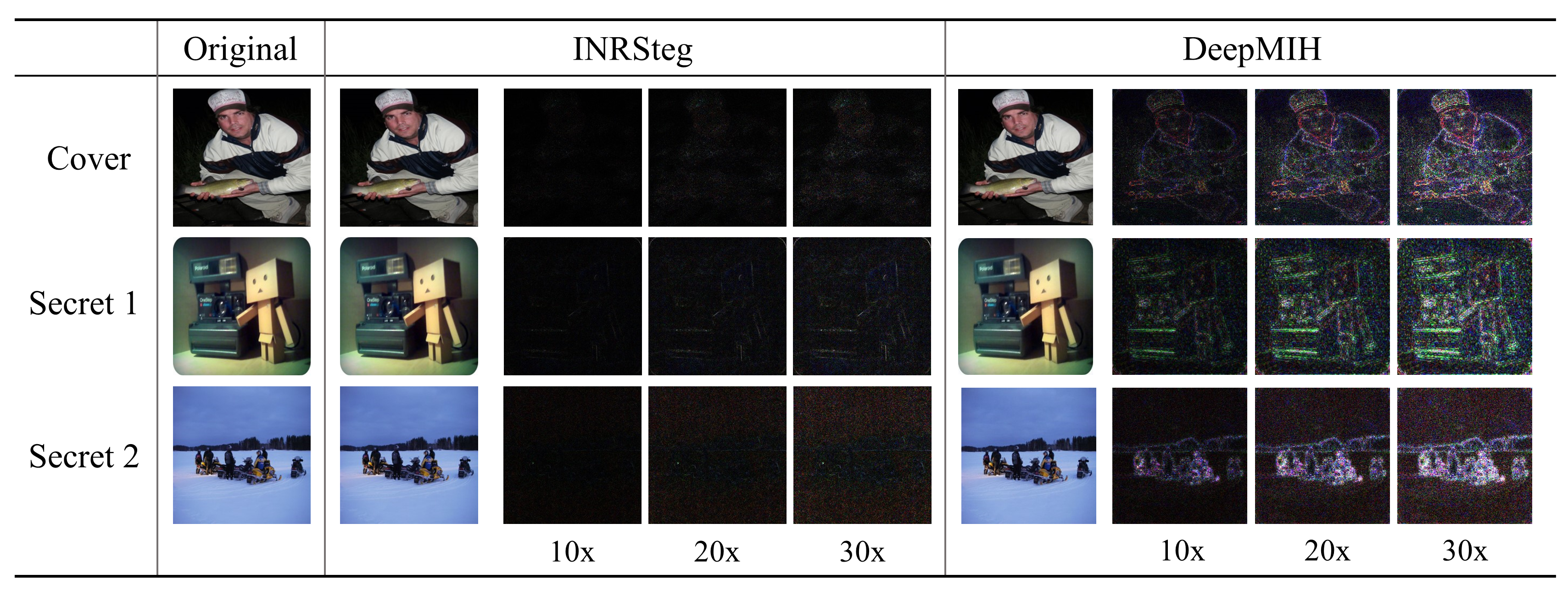}
  \caption{A comparison of difference maps (enhanced by 10x, 20x, and 30x) between the original images and the revealed images. Columns 2-6 are results of INRSteg and columns 7-10 are results of DeepMIH \cite{9676416}.
  }
  \label{fig: image difference map}
\end{figure}

\subsubsection{Image steganalysis}
To examine security, we adopt two image steganalysis tools, SiaStegNet \cite{9153041} and XuNet \cite{xu2017deep}. \cref{table: image-to-image steganalysis} presents the detection accuracy of our framework and other models. The security is higher as the detection accuracy is closer to 50\%. Unlike other steganography models, the detection accuracy of INRSteg is 50$\%$, meaning that the steganalysis tool completely fails to distinguish the stego data from the cover data. This shows that our framework is undetectable using existing steganalysis tools and assures perfect security. This is because our framework does not directly edit the data in the discrete representation state, but performs manipulation after transforming the representation into a neural network, which is unnoticeable when re-transformed to the discrete representation. Moreover, we conduct steganalysis analysis on multi-images steganography. The results in Appendix F.

\subsubsection{Multi-image steganography}
In \cref{table: multi-image steganography}, we conduct experiments for multi-image steganography hiding more than two images when the size of stego INR is fixed, so that one may have to decrease the network size of each hidden data in order to fit the memory capacity. The results show that high recovery performance maintains even when increasing the number of hidden data. For more experiments with other modalities, please refer to Appendix G.

\noindent
\begin{minipage}[h]{0.59\textwidth}
\begin{table}[H]
\centering
\captionsetup{width=0.9\textwidth}
\caption{Performance comparison of image-to-image steganography.}
\label{table: image-to-image steganography comparison}
\scriptsize
\renewcommand{\arraystretch}{1.1}
\setlength{\tabcolsep}{2.5pt}
\begin{tabular}{c|c|cccc}
\toprule
& Metric & MAE $\downarrow$  & PSNR $\uparrow$ &  SSIM $\uparrow$ & RMSE $\downarrow$ \\
\midrule
\multirow{2}{*}{DeepSteg} & cover & 2.576 & 37.48 & 0.947 & 3.417\\
& secret & 3.570 & 33.34 & 0.957 & 3.57\\
\midrule
\multirow{2}{*}{DeepMIH} & cover & 2.839 & 36.05 & 0.932 & 4.276 \\
& secret & 3.604 & 32.97 & 0.9550 & 5.746\\
\midrule
\multirow{2}{*}{INRSteg} & cover & \textbf{0.153} & \textbf{62.34} & \textbf{0.999} & \textbf{0.204}\\
& secret & \textbf{1.084}& \textbf{45.84} & \textbf{0.988} & \textbf{1.326}\\
\bottomrule
\end{tabular}
\end{table}
\end{minipage}
\begin{minipage}[h]{0.39\textwidth}
\renewcommand{\arraystretch}{1.05}
\begin{table}[H]
\centering
\captionsetup{width=0.95\textwidth}
\caption{Steganalysis comparison of image-to-image steganography.}
\label{table: image-to-image steganalysis}
\scriptsize
\renewcommand{\arraystretch}{1.135}
\setlength{\tabcolsep}{4.5pt}
\begin{tabular}{c|cc}
\toprule
\multirow{2}{*}{} & \multicolumn{2}{c}{Accuracy (\%) } \\
\cmidrule(rl){2-3}
& SiaStegNet & XuNet \\
\midrule
{ DeepSteg } & 92.87 & 75.83\\
\midrule
{ DeepMIH } & 90.70 & 90.82\\
\midrule
{ INRSteg } & \textbf{50.00} & \textbf{50.00} \\
\bottomrule
\end{tabular}
\end{table}
\end{minipage}

\subsection{Efficiency and Robustness Analysis}
\label{experiment: Efficiency and Robustness Analysis}

\subsubsection{Model Size Comparison}
\label{experiment: model size comparison}
In our efficiency analysis, we compare model sizes to evaluate the computational cost and efficiency of our proposed INRSteg relative to existing steganography models. As described in \cref{table:model_params}, our INRSteg achieves a remarkably compact architecture, consisting of only 400,000 parameters. Compared to other existing steganography models, this advantage is notable. DeepSteg\cite{NIPS2017_838e8afb}, with 42.58 million parameters, is approximately 107.46 times larger than our model. DeepMIH\cite{9676416} model, with 12.42 million parameters, has 31.34 times more parameters, and LF-VSN\cite{mou2023large}, at 7.40 million parameters, is 18.68 times larger. Especially, CRoSS\cite{yu2024cross} contains 1,066.24 million parameters, corresponding to 2690.54 times the size of the model. This comparative analysis highlights the efficiency of our INRSteg, demonstrating its capability to achieve robust steganography performance with a significantly reduced computational space. The small model not only promotes faster training and inference times, but also opens up opportunities for deployment on devices with limited computational resources such as smartphones.

\noindent
\begin{minipage}[t]{0.5\textwidth}
\begin{table}[H]
\centering
\captionsetup{width=0.9\textwidth}
\caption{Distortion performance of hiding multiple images ($>$ 2) into one.}
\label{table: multi-image steganography}
\scriptsize
  \renewcommand{\arraystretch}{1.15}%
  \setlength{\tabcolsep}{3pt}
\begin{tabular}{c|c|cccc}
\toprule
&Metric & PSNR $\uparrow$ & RMSE $\downarrow$ \\
\midrule
\multirow{2}{*}{3 images} & cover & 62.31 & 0.204\\
&secret 1,2,3 & 43.67& 1.642\\
\midrule
\multirow{2}{*}{4 images} & cover & 62.58 & 0.153\\
&secret 1,2,3,4 & 41.86 & 2.059\\
\bottomrule
\end{tabular}
\end{table}
\end{minipage}
\begin{minipage}[t]{0.5\textwidth}
\begin{table}[H]
\centering
\caption{Model Size Comparison.}
\label{table:model_params}
\scriptsize
  \renewcommand{\arraystretch}{1.1} 
  \setlength{\tabcolsep}{3pt}
\begin{tabular}{c|cc}
\toprule
{Model} & {\#Params} & {Ratio to Ours} \\
\midrule
DeepSteg \cite{NIPS2017_838e8afb} & 42.58M & 107.46${\times}$ \\
HiNet \cite{Jing_2021_ICCV} & 4.05M & 10.22${\times}$ \\
DeepMIH \cite{9676416} & 12.42M & 31.34${\times}$ \\
CRoSS \cite{yu2024cross} & 1,066.24M & 2690.54${\times}$ \\
LF-VSN \cite{mou2023large} & 7.40M & 18.68${\times}$ \\
Ours & 0.40M & 1${\times}$ \\
\bottomrule
\end{tabular}
\end{table}
\end{minipage}

\subsubsection{Robustness}
\label{experiment: Robustness} 
We explore the robustness of INRSteg under harsh conditions in real-world scenarios. In particular, we apply quantization operation to store model parameters from float32 to int8 format. This process aims to simulate scenarios where data precision must be reduced due to storage or transmission constraints. After quantization, we recover the parameters from int8 to float32 and evaluate the performance metrics of the recovered secret images. \cref{table:quantization_performance} describes the results of performance comparison before and after quantization across three cover types: image, audio, and video. Notably, while PSNR values decrease after quantization, this decrease is not as significant as might be expected given the harsh nature of the quantization process. This indicates that INRSteg has a robust ability to maintain secret data quality despite the data compression and precision reductions. Furthermore, something to crucially note in \cref{table:quantization_performance} is that the SSIM value hardly changes after quantization. Since SSIM measures the perceptual quality, it is a more representative metric that evaluates image quality. Therefore, the results indicate that the visual content of the secret images remain almost completely. The reconstruction results are in Appendix H. Additionally, we show preserving secret data when using another model compression method, pruning, in Appendix I.

\begin{table}[h!]
\centering
\caption{Performance Metrics Before and After Quantization for Different Cover Types. Pre-Q denotes metrics before quantization, Post-Q denotes metrics after quantization. Higher values are better for PSNR and SSIM, while lower values are preferred for RMSE and MAE.}
\label{table:quantization_performance}
\scriptsize
  \renewcommand{\arraystretch}{1.1}%
  \setlength{\tabcolsep}{4pt} 
\begin{tabular}{c|cccccccc}
\toprule
 & \multicolumn{2}{c}{PSNR $\uparrow$} & \multicolumn{2}{c}{RMSE $\downarrow$} & \multicolumn{2}{c}{SSIM $\uparrow$} & \multicolumn{2}{c}{MAE $\downarrow$} \\
\cmidrule{2-9}
Cover type & { Pre-Q} & {Post-Q } & { Pre-Q} & {Post-Q } & { Pre-Q} & {Post-Q } & { Pre-Q} & {Post-Q } \\
\midrule
Image & 40.0766 & 33.7483 & 2.5245 & 5.2375 & 0.9821 & 0.9554 & 1.7085 & 3.519 \\
Audio & 40.0766 & 33.5438 & 2.5245 & 5.3623 & 0.9821 & 0.9529 & 1.7085 & 3.723 \\
Video & 40.0766 & 33.8276 & 2.5245 & 5.1899 & 0.9821 & 0.9583 & 1.7085 & 3.417 \\
\bottomrule
\end{tabular}
\end{table}

\section{Limitations and Discussion}
\label{Limitations and Discussion}
We would like to highlight the potential negative impacts that steganography frameworks may result in the ethical aspect. Users must be cautious not to utilize our work in unethical situations, where one may hide inappropriate information. Additional limitations are described in Appendix J.

\section{Conclusion}
\label{conclusion}
In conclusion, this paper introduces INRSteg, an innovative framework designed for the concealment of multiple cross-modal data through the utilization of the weight space inherent in Implicit Neural Representations (INR). Our extensive experimentation demonstrates that INRSteg surpasses existing steganography techniques in key areas, namely distortion evaluation, data capacity, and security measures, across both intra and cross-modal steganography scenarios. A notable advantage of INRSteg is its deployment of smaller model sizes, which directly contributes to reduced energy consumption, making it exceptionally suited for practical real-world applications. Furthermore, our analysis confirms the robustness of INRSteg, even when subjected to the rigorous demands of quantization operations. We believe that INRSteg establishes a new benchmark for achieving an optimal balance between efficiency and performance within the steganography domain, paving the way for future advancements in secure and efficient steganography techniques.

\section*{Acknowledgements}
\label{acknowledgements}
This work was partly supported by Institute of Information \& communications Technology Planning \& Evaluation (IITP) grant funded by the Korea government(MSIT) (No.RS-2022-II220184, Development and Study of AI Technologies to Inexpensively Conform to Evolving Policy on Ethics) and Institute for Information \& communications Technology Planning \& Evaluation (IITP) grant funded by the Korea government(MSIT) (No. 2021-0-01381, Development of Causal AI through Video Understanding and Reinforcement Learning, and Its Applications to Real Environments).

\par\vfill\par

\clearpage

\bibliographystyle{splncs04}
\bibliography{main}

\end{document}


\title{Supplementary Materials Template: }

\author{Sojeong Song* \and Seoyun Yang* \and Chang D. Yoo \and Junmo Kim\\
School of Korea Advanced Institute of Science and Technology (KAIST)\\
Daejeon, South Korea\\
{\tt\small akqjq4985@kaist.ac.kr \and seoyun.yang@navercorp.com}
}


\section{Supplementary A}
\label{experiment: experimental setting}

\textbf{Dataset Setup} For ImageNet, each image data is resized to 256 x 256 resolution. We further utilize mini-ImageNet for steganalysis experiments. For GTZAN music genre dataset, each audio data is cropped to a length of 100,000 samples with a 22,050 sample rate. For the scikit-video dataset, each video data is resized to 128 × 128 resolution and 16 frames are extracted. From the Stanford 3D Scanning Repository, we select two samples, the Asian Dragon and the Armadillo. VoxCeleb2 is a large scale audio-visual dataset, so that it is widely used for a number of various applications such as visual speech synthesis, cross-modal transfer task. For VoxCeleb2 dataset, we sample 100 pairs from test set and each video data and audio data undergo same preprocessing like GTZAN music genre dataset and scikit-video dataset.

\textbf{Implementation Details} For INR, we use SIREN as the base structure. In this paper, we refer to the SIREN structure for cover INRs. For single-data steganography, the neural networks are padded with a ratio of $1.0$. For steganalysis experiments using images, we use 5 hidden layers with a size of 128. The Adam optimizer with learning rate 1e-4 is used for all modalities.

\textbf{Distortion evaluation metric} To evaluate the distortion between cover and reconstructed cover data, and secret and revealed secret data, we visualize 3D shapes and employ several evaluation metrics for image, audio, and video. For image data, we employ Peak Signal-to-Noise Ratio (PSNR), Structural Similarity Index (SSIM), Mean Absolute Error (MAE), and Root Mean Square Error (RMSE). For video data, in addition to PSNR and SSIM we evaluate Average Pixel Discrepancy (APD) and Perceptual Similarity (LPIPS). For audio data, we employ MAE and Signal-to-Noise Ratio (SNR). All data are represented within the range of 0 to 255.

\section{Supplementary B}
\label{appendix: padding}
\begin{figure}[!h]
\begin{center}
\centerline{\includegraphics[width=\columnwidth, height=\textwidth]{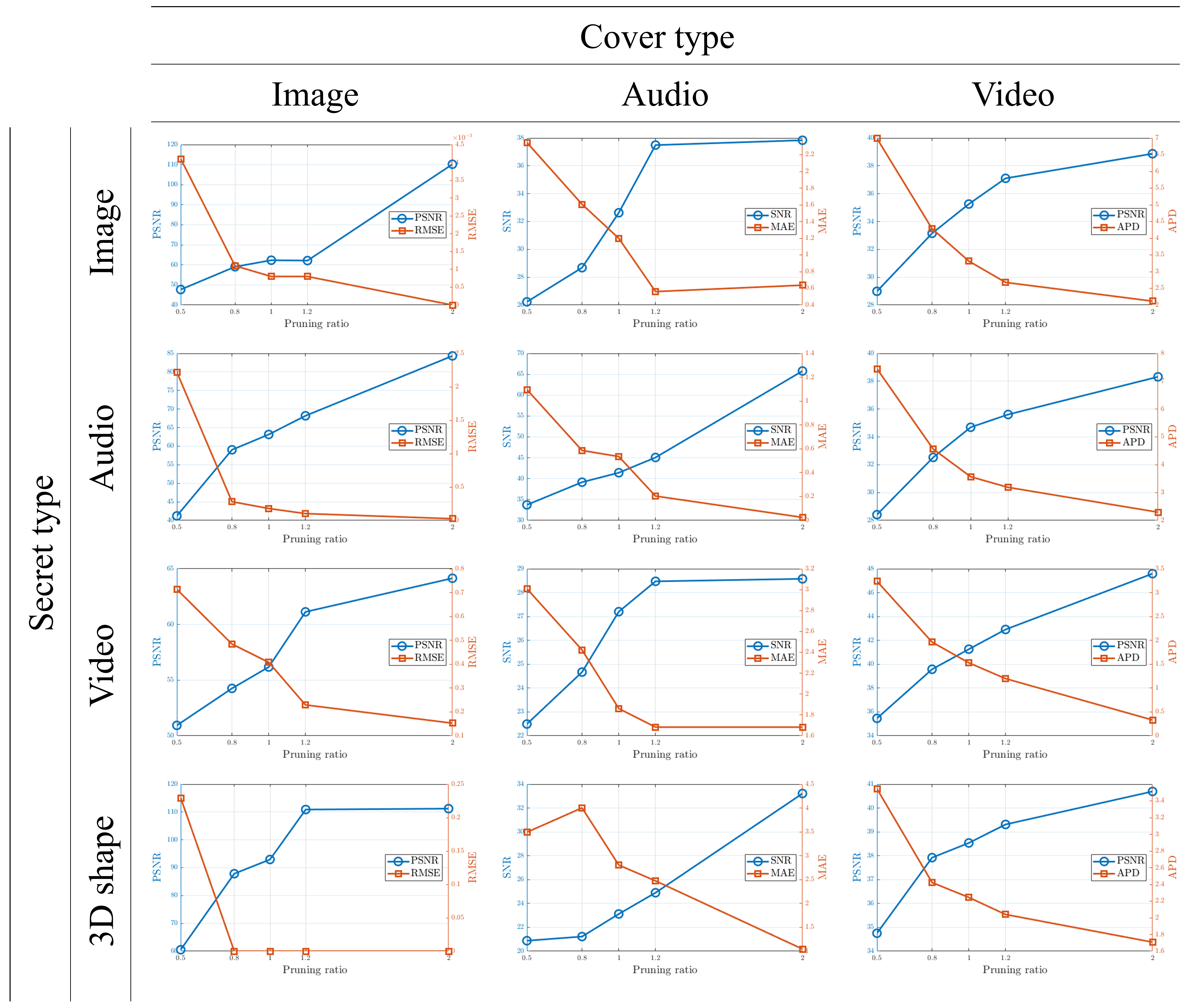}}
\caption{Ablation experiments for hiding single-data steganography varying padding rate. Padding rate refers to the division of padded amount on secret INR by secret INR's hidden layer size. The padding rate varies across ${[0.5, 0.8, 1.0, 1.2, 2.0]}$. The evaluation metrics are PSNR/RMSE, SNR/MAE, and PSNR/APD for image, audio, and video, respectively. Early stopping is executed in this experiment.}
\label{fig: padding}
\end{center}
\end{figure}

Ablation study is performed to figure out the effectiveness of the padding rate in single-data steganography. In this experiment, we set the size of hidden layer at 128, 128, 192, and 192 for image, audio, video, and 3d shapes, respectively. The INR architecture for secret data is fixed and that of stego INR depends on the cover data type and padding rate. \cref{fig: padding} shows the comparison between cover data and stego data which is measured by various evaluation metrics. As shown in \cref{fig: padding}, as the padding rate is bigger, which means that the space to be fitted on cover data is larger, the network overfits well on cover data, resulting in stego data with high similarities with the cover data. Interestingly, it can be seen that only adding 50${\%}$ of the network is enough to get high PSNR, which means our method is very efficient in making stego INR. This may be due to the good weight initialization from secret data. Moreover, if the secret and cover data have the same modality, fitting is well performed.
\noindent

\section{Supplementary C}
\label{appendix: reconstruction results}
\begin{figure}[ht]
    \centering
    \begin{minipage}{.5\textwidth}
        \centering
        \includegraphics[width=\linewidth]{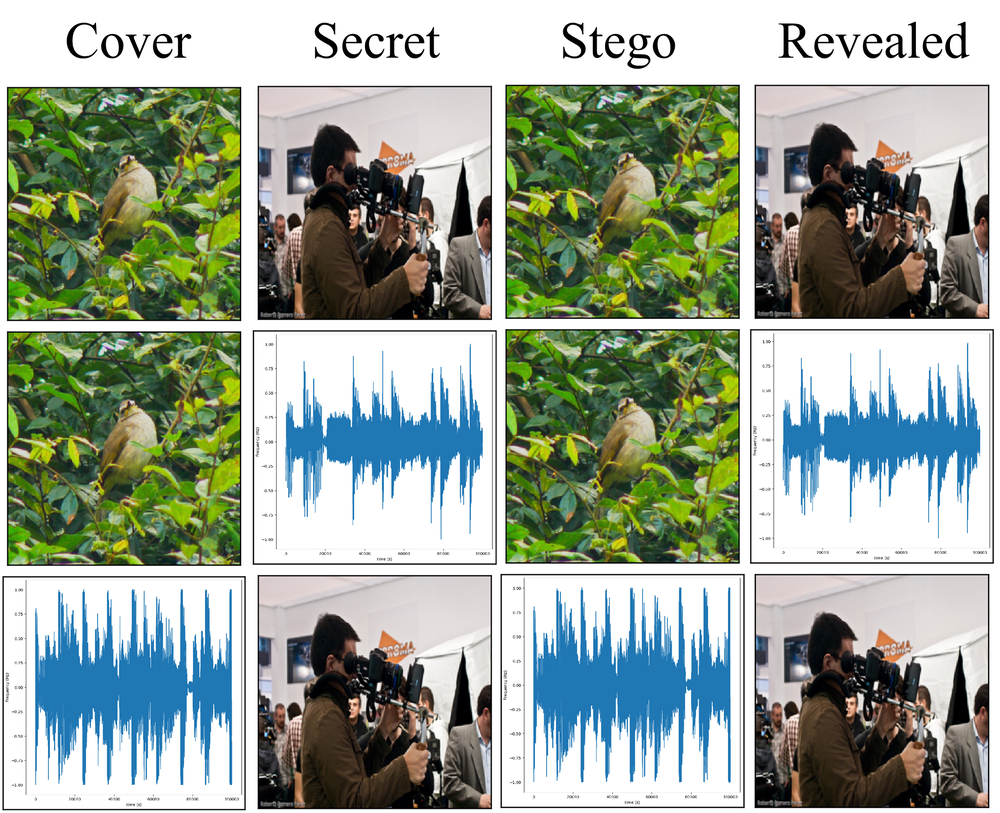}
        \captionsetup{width=0.95\textwidth}
        \caption{image-to-image, audio-to-image, and image-to-audio steganography in order from the top to the bottom.}
        \label{fig: recon onetoone}
    \end{minipage}%
    \begin{minipage}{.45\textwidth}
        \centering
        \includegraphics[width=\linewidth]{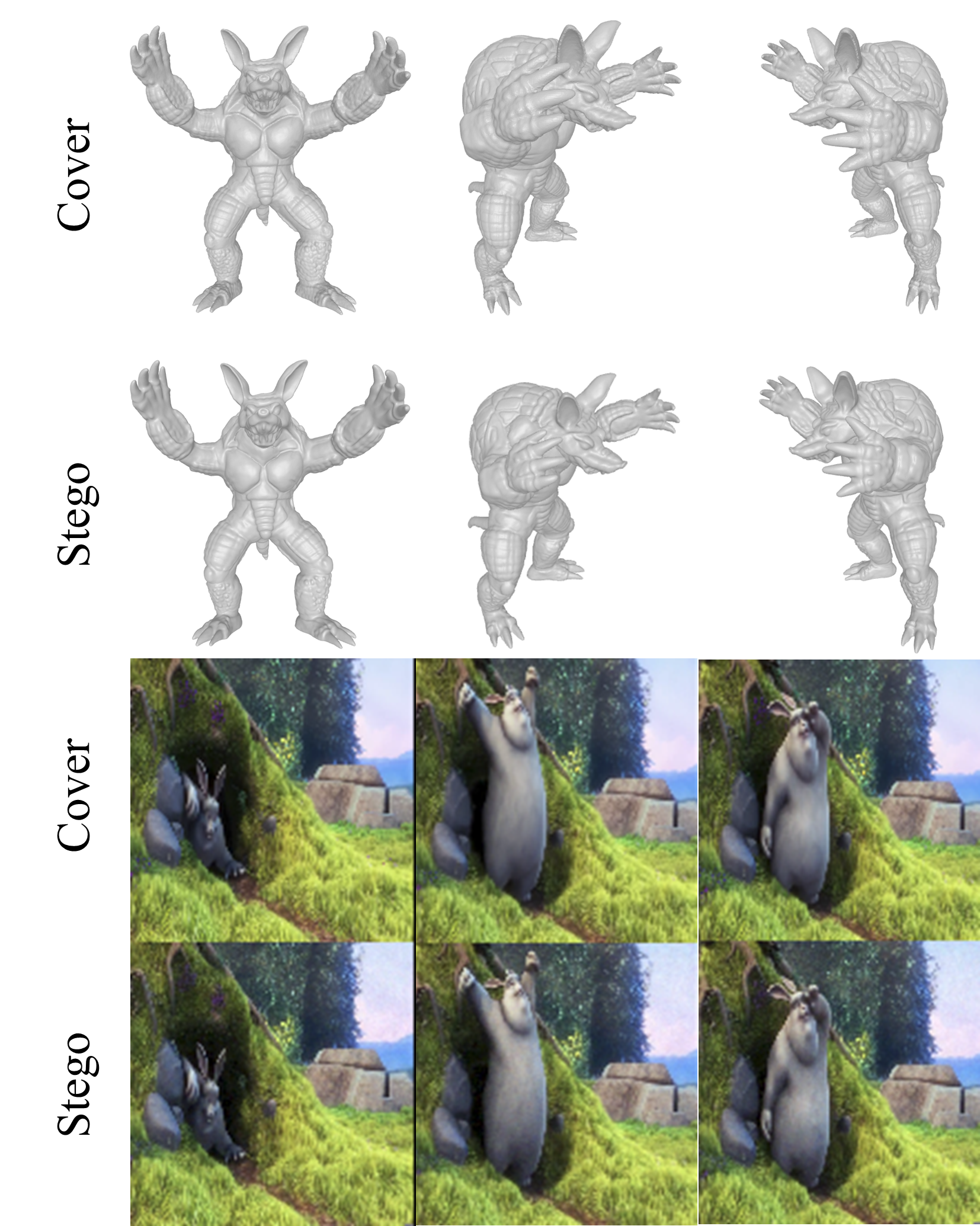}
        \captionsetup{width=0.98\textwidth}
        \caption{Stego/Cover visualization for multi-data steganography: 3D shape and audio to 3D shape, and image and 3D shape to video.}
        \label{fig: recon 3d video}
    \end{minipage}
\end{figure}

\begin{figure}[ht]
\begin{center}
\centerline{\includegraphics[width=0.7\columnwidth]{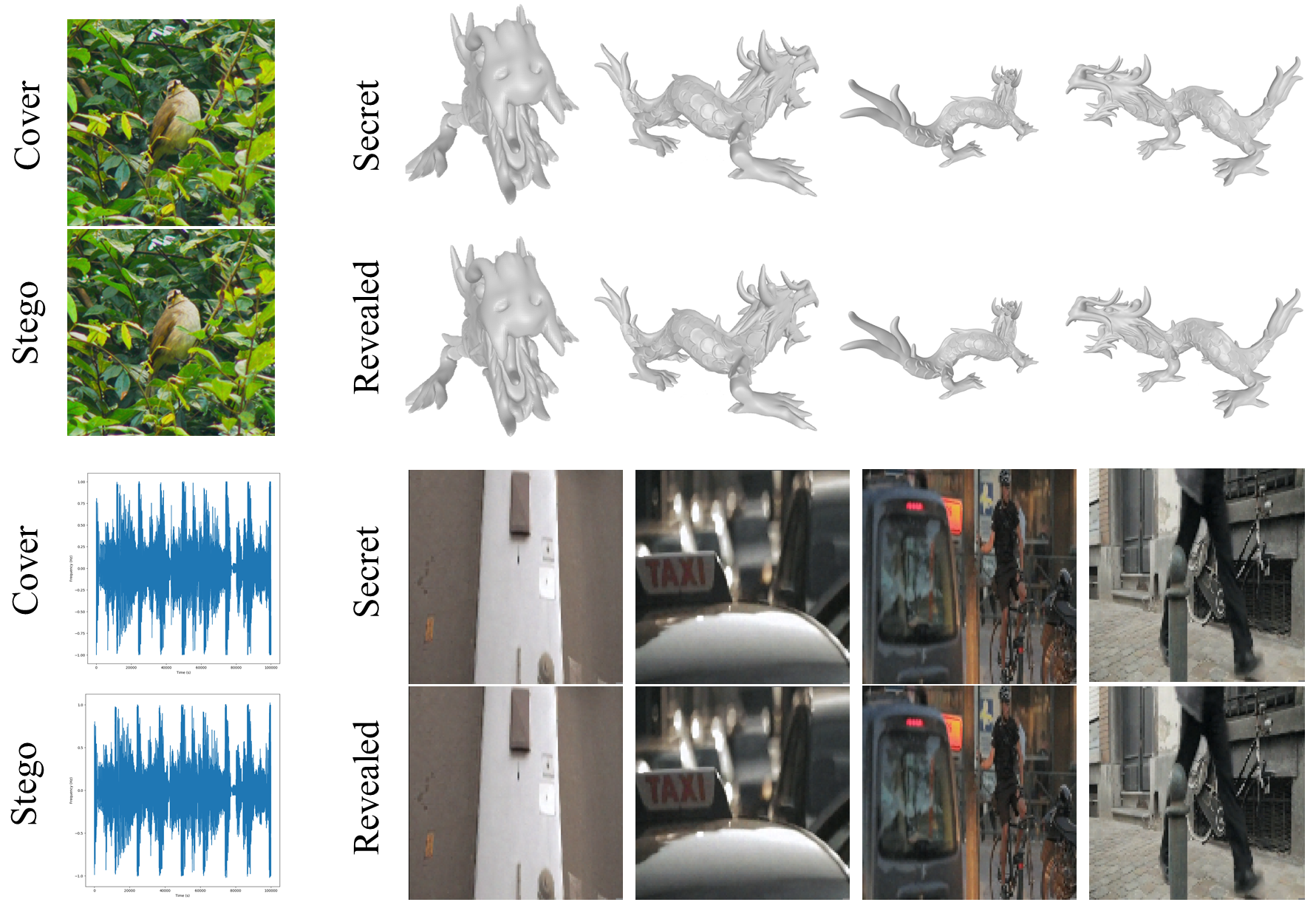}}
\caption{3D shape-to-image and video-to-audio steganography in order from the top to the bottom.}
\label{fig: recon twotoone}
\end{center}
\end{figure}

In this appendix, we provide reconstruction results of various steganography experiments, focusing on the reconstruction quality. \cref{fig: recon onetoone} visualizes the reconstructed images and waveform of audio for image-to-image, audio-to-image, and image-to-audio steganography. Comparing cover data with the reconstructed data from stego INR, it reveals that there exists no difference between them for both images and audio. For secret and revealed secret data, they are visually same, which indicates the effectiveness of INRSteg in achieving successful cross-modal steganography. Similarly, \cref{fig: recon twotoone} presents successful reconstruction results for 3D shape-to-image and video-to-audio steganography demonstrating adaptability of our INRSteg. Additional results on cross-modal multi-data steganography are presented in \cref{table: multi-data cross steganography} and are visually supported by \cref{fig: recon twotoone}, underlining the robust performance of our method across various data and modality combinations.

\begin{table}[H]
\centering
\caption{Performance of three steganography tasks: audio and video to audio, image and 3D shape to video, and video and 3D shape to image. Metrics for 3D shapes are excluded as they are visualized in \cref{fig: recon twotoone}.}
\label{table: multi-data cross steganography}
\footnotesize
\setlength{\tabcolsep}{2.3pt}
\renewcommand{\arraystretch}{0.9}
\begin{tabular}{p{2.5cm}|p{2.2cm}p{2.2cm}p{2.2cm}p{2.2cm}}
\toprule
\multicolumn{1}{c|}{Metric}  & \hfil MAE/APD $\downarrow$  & \hfil SNR/PSNR $\uparrow$ & \hfil SSIM $\uparrow$ & \hfil LPIPS $\downarrow$ \\
\midrule
\hfil Cover (audio) & \hfil 0.497 & \hfil 30.91 & \hfil - & \hfil -\\
\hfil Secret 1 (audio) & \hfil 0.854 & \hfil 25.70 & \hfil - & \hfil - \\
\hfil Secret 2 (video) & \hfil 2.907 & \hfil 36.64 & \hfil 0.997 & \hfil 0.032\\
\midrule
\hfil Cover (video) & \hfil 3.366 & \hfil 35.38 & \hfil 0.997 & \hfil 0.060\\
\hfil Secret 1 (image) & \hfil 1.084& \hfil 45.84 & \hfil 0.988 & \hfil 1.326 \\
\midrule
\hfil Cover (image)& \hfil 0.740 & \hfil 48.58 & \hfil 0.993 & \hfil 0.944\\
\hfil Secret 1 (video) & \hfil 2.907 & \hfil 36.64 & \hfil 0.997 & \hfil 0.032\\
\bottomrule
\end{tabular}
\end{table}

\noindent

\noindent

\section{Supplementary D}
\label{appendix: permutation}
\begin{figure}[ht]
    \centering
    \begin{minipage}{.5\textwidth}
        \centering
        \includegraphics[width=\linewidth]{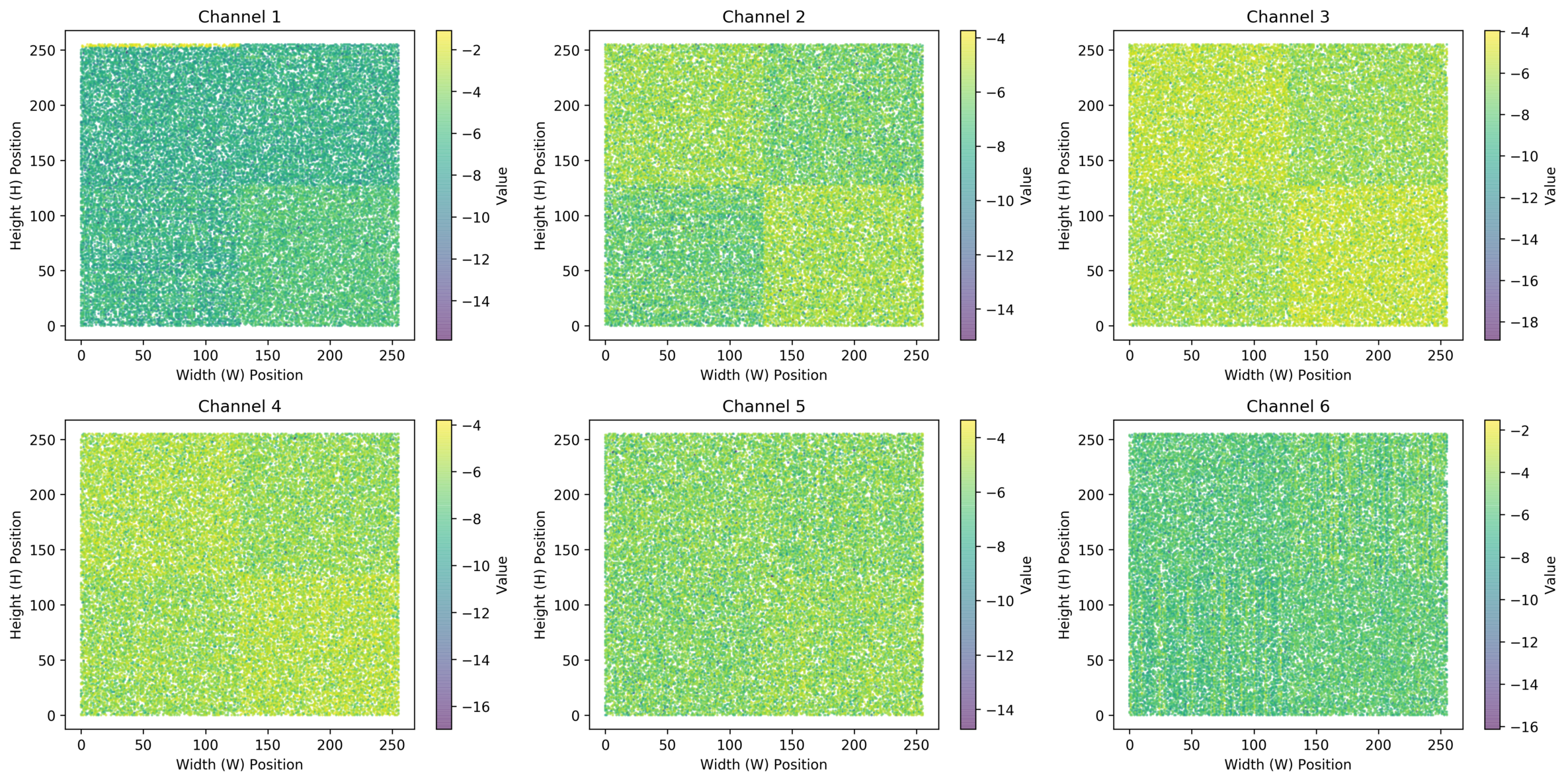}
        \captionsetup{width=0.95\textwidth}
        \caption{Visualization of the weight distribution of a stego INR \textbf{before} permutation. Natural logarithm is applied to magnify the weight difference.}
        \label{fig:pre_perm}
    \end{minipage}%
    \begin{minipage}{.5\textwidth}
        \centering
        \includegraphics[width=\linewidth]{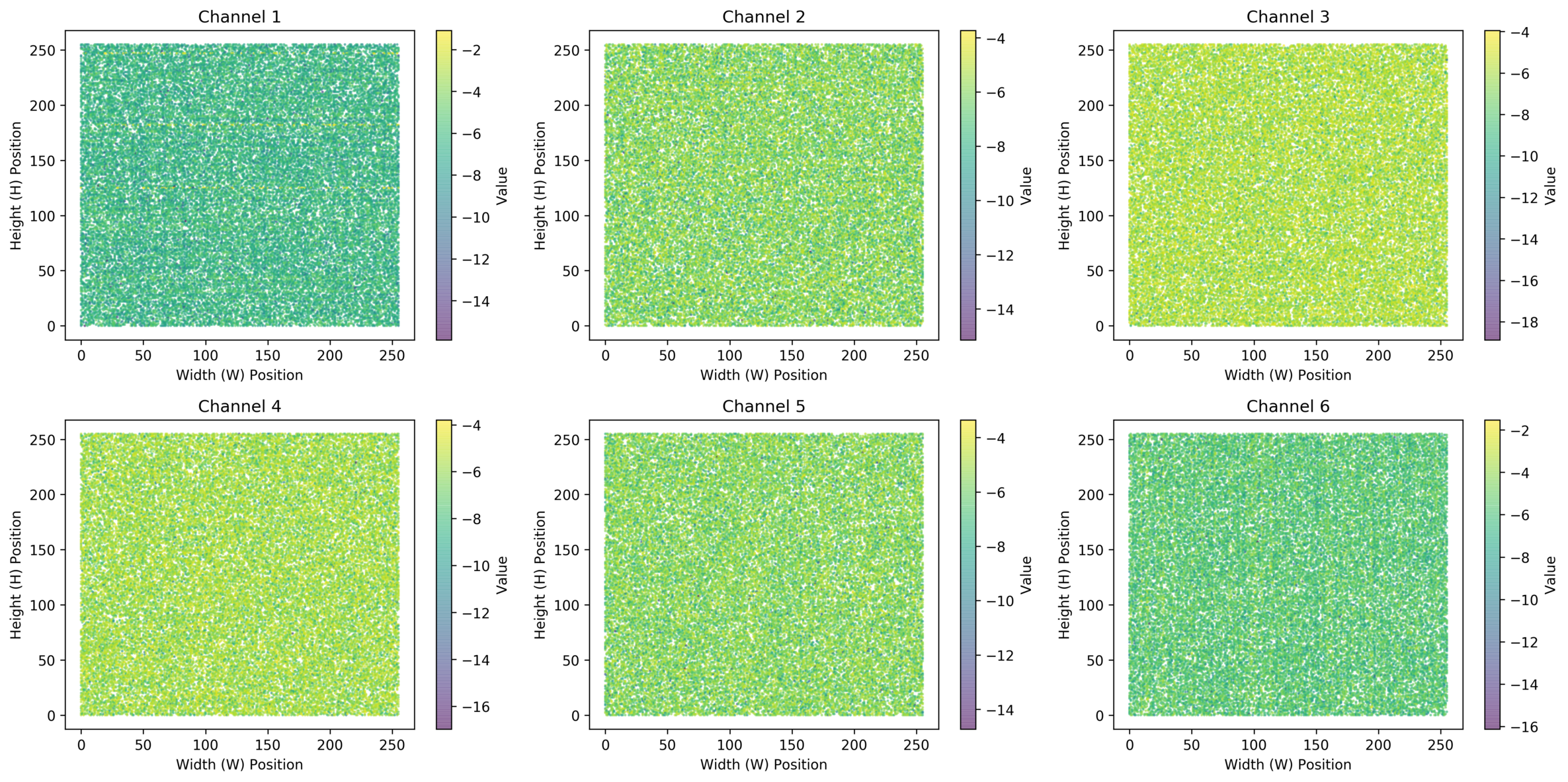}
        \captionsetup{width=0.95\textwidth}
        \caption{Visualization of the weight distribution of a stego INR \textbf{after} permutation. Natural logarithm is applied to magnify the weight difference.}
        \label{fig:post_perm}
    \end{minipage}
\end{figure}

\begin{figure}[ht]
\begin{center}
\centerline{\includegraphics[width=\columnwidth]{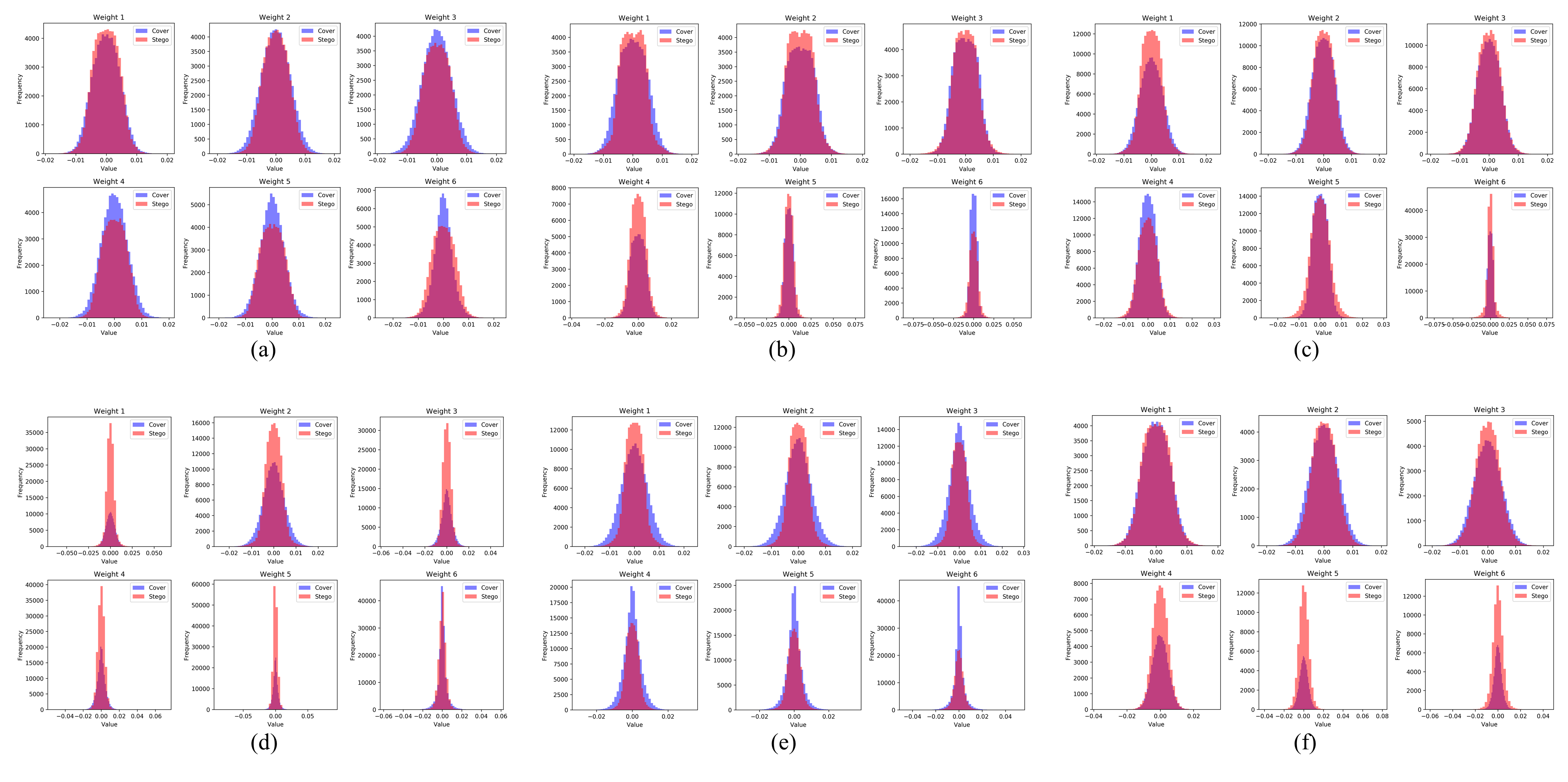}}
\caption{Six examples of overlapped weight histogram. Blue one is weight histogram of INR which does not contain any secret data while red one is weight histogram of stego INR which contain secret data inside. We randomly select six examples covering a variety cases from (a) to (f).}
\label{fig: weight distribution}
\end{center}
\end{figure}

In this appendix, we perform security analysis of INRSteg by visualizing weight space and compare weight distribution. Implicit Neural Representations, unlike discrete representations, is not definite, so the distribution of the weight space differs for each data regardless of the modality and the appearance. Therefore, the weight space of the stego INR contains borderlines that indicate the positions of the hidden data, which can be shown in \cref{fig:pre_perm}. To overcome this problem, we propose permutation via private key to ensure security even in the aspect of INRs. After the neural network of the stego data goes through permutation, the weight space becomes indistinguishable as in \cref{fig:post_perm}, therefore preventing potential steganalysis algorithms from detecting the existence of secret data. Additionally, through permutation, we can effectively encode the location of the secret data, so that the secret data cannot be revealed without the private key used for permutation.  
\noindent
We further explore the indistinguishability between secret and cover data within the weight space by arranging the weights in order of magnitude and visualizing their distribution through weight histograms. \cref{fig: weight distribution} shows comparison between the histograms of stego INRs and those of INRs which trained without any embedded secret data. The stego INR appears a singular INR, which can be seen in red weight histograms of stego INR. Through overlapped histograms, we observe that there exists no visual criterion that distinguishes the stego INR and the INR which does not contain secret data. This observation validates the security of INRSteg.
\noindent

\section{Supplementary E}
\label{appendix: cross-modal single-data secret/revealed performance}

For ablation study of cross-modal single-data steganography, \cref{tab:cross-modal single-data secret/revealed} shows secret and revealed secret performance of all cross-modal cases. In this ablation study, the size of secret INR is fixed for all cases using padding rate of 1.0, we set the size of hidden layer at 128, 128, 192, and 192 for image, audio, video, and 3d shapes, respectively. For the secret and the revealed secret data pair, as our framework goes through lossless recovery in terms of INR, the secret/revealed secret performance is solely dependent on the data representation transformation phase. This performance can either be increased or decreased flexibly as one of the strengths of our framework is that we can control the secret/revealed secret performance by manipulating the network size for the secret data.

\begin{table}[!h]
  \centering
  \caption{Secret/Revealed secret performance of all steganography tasks.}
  \label{tab:cross-modal single-data secret/revealed}
  \begin{tabular}{@{}>{\centering\arraybackslash}m{1.9cm}>{\centering\arraybackslash}m{1.9cm}|>{\centering\arraybackslash}m{1.9cm}>{\centering\arraybackslash}m{1.9cm}|>{\centering\arraybackslash}m{1.9cm}>{\centering\arraybackslash}m{1.9cm}@{}}
    \toprule
    \multicolumn{6}{c}{{{Secret}}} \\
    \midrule
    \multicolumn{2}{c|}{{image}} & \multicolumn{2}{c|}{{audio}} & \multicolumn{2}{c}{{video}} \\
    \midrule
    PSNR $\uparrow$ & RMSE $\downarrow$ & SNR $\uparrow$ & MAE $\downarrow$ & PSNR $\uparrow$ & APD $\downarrow$ \\
    45.84 & 1.326 & 25.70 & 0.854 & 36.64 & 2.907 \\
    \bottomrule
  \end{tabular}
\end{table}

\section{Supplementary F}
\label{appendix: steganalysis}
In our manuscript, we illustrate how INRSteg completely avoids detection through steganalysis techniques in the image-to-image steganography task. This section extends our experiment, presenting comparison results between DeepMIH and our approach, in the case of multi-image steganography. Similar to the results of single-image steganography, \cref{table: multi intra-modal steganography comparison} and \cref{table: Steganalysis} show that INRSteg significantly outperforms ensuring successful steganography at a higher quality and achieving perfect security when using steganalysis models, SiaStegNet and XuNet.

\noindent
\begin{minipage}[t]{0.63\textwidth}
\begin{table}[H]
\centering
\captionsetup{width=0.9\textwidth}
\caption{Performance comparison of two images into one image steganography.}
\label{table: multi intra-modal steganography comparison}
\footnotesize
\renewcommand{\arraystretch}{1.0}
\setlength{\tabcolsep}{0.5pt}
\begin{tabular}{c|c|cccc}
\toprule
& Metric & {MAE$\downarrow$ }  & {PSNR$\uparrow$ } &  {SSIM$\uparrow$ } & {RMSE$\downarrow$} \\
\midrule
\multirow{3}{*}{DeepMIH} & cover & 2.754 & 35.78 & 0.941 & 4.318\\
& secret 1 & 3.604 & 32.97 & 0.9550 & 5.746\\
& secret 2 & 2.6095 & 36.34 & 0.959 & 3.919\\
\midrule
\multirow{2}{*}{INRSteg} & cover & \textbf{0.204} & \textbf{62.42} & \textbf{0.999} & \textbf{0.255}\\
& { secret 1, 2 } & \textbf{1.084}& \textbf{45.84} & \textbf{0.988} & \textbf{1.326}\\
\bottomrule
\end{tabular}
\end{table}
\end{minipage}
\begin{minipage}[t]{0.335\textwidth}
\begin{table}[H]
\centering
\caption{Steganalysis comparison of two images into one image steganography.}
\label{table: Steganalysis}
\footnotesize
\renewcommand{\arraystretch}{1.12}
\begin{tabular}{c|cc}
\toprule
\multirow{2}{*}{} & \multicolumn{2}{c}{Accuracy (\%) } \\
\cmidrule(rl){2-3}
& SiaStegNet & XuNet \\
\midrule
DeepMIH & 87.96 & 68.52\\
\midrule
INRSteg & \textbf{50.00} & \textbf{50.00} \\
\bottomrule
\end{tabular}
\end{table}
\end{minipage}

\section{Supplementary G}
\begin{wraptable}{r}{6cm}
\footnotesize
\centering
\captionsetup{width=0.45\textwidth}
\caption{Distortion performance of hiding multiple audios ($>$ 2) into one when the secret INR hidden layer width (D) is set to 86 and 64 for hiding 3 audios and 4 audios respectively.}
\label{table: multi-audio steganography}
\begin{tabular}{c|c|cccc}
\toprule
&Metric & SNR $\uparrow$ & MAE $\downarrow$ \\
\midrule
\multirow{2}{*}{3 audios} & cover & 41.45 & 0.232\\
&secret 1,2,3 & 17.98 & 2.222\\
\midrule
\multirow{2}{*}{4 audios} & cover & 42.56 & 0.153\\
&{ secret 1,2,3,4 } & 14.81 & 2.860\\
\bottomrule
\end{tabular}
\end{wraptable}To evaluate adaptability of our INRSteg across modalities beyond images, we conduct additional experiment hiding more than two audio into one. The size of stego INR is fixed and \cref{table: multi-audio steganography} presents distortion performance metrics. As can be seen in \cref{table: multi-audio steganography}, cover maintains a high SNR with a low MAE, whereas those for the secret audios exhibit a lower SNR and a higher MAE. This performance degradation correlates with the reduction in the secret network size, showing a trade-off between the number of audios embedded and the quality of the secret and revealed secret audios.

\section{Supplementary H}
\begin{figure}[!h]
\begin{center}
\centerline{\includegraphics[width=0.8\columnwidth]{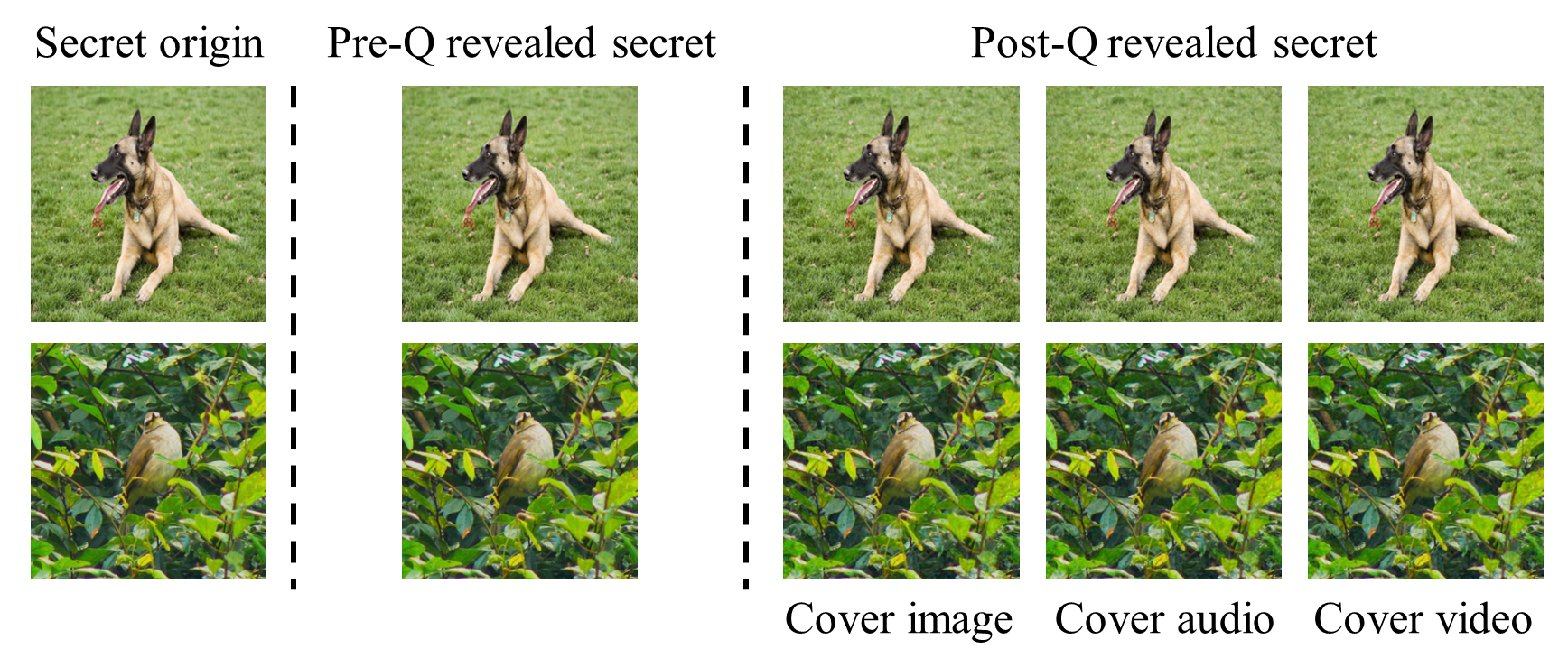}}
\caption{Visualization of revealed secret data before and after quantization operation. 'Pre-Q Revealed Secret' denotes revealed secret data from stego INR which is not applied quantization. 'Post-Q Revealed Secret' denotes revealed secret data from damaged stego INR after quantization and dequantization.}
\label{fig: recon quantization}
\end{center}
\end{figure}

In the main paper, we explore the robustness of INRSteg by applying quantization on the INR. For implementation details of quantization operation, we calculate the scale ${s}$ and zero-point ${z}$ based by using the range of weight values and that of target bit precision. Weight values are converted to int8 format by using ${torch.clamp}$ and ${torch.round}$ function by using ${s}$ and ${z}$ values. Subsequently dequantization is applied to the weight values by using scale and zero-point values. In this section, we visualize the reconstruction results to demonstrate that  the visual content of the secret images remain almost completely. \cref{fig: recon quantization} shows no visual degradation of quality for all revealed secret data.
\noindent

\section{Supplementary I}
\begin{figure}[ht]
\begin{center}
\centerline{\includegraphics[width=\columnwidth]{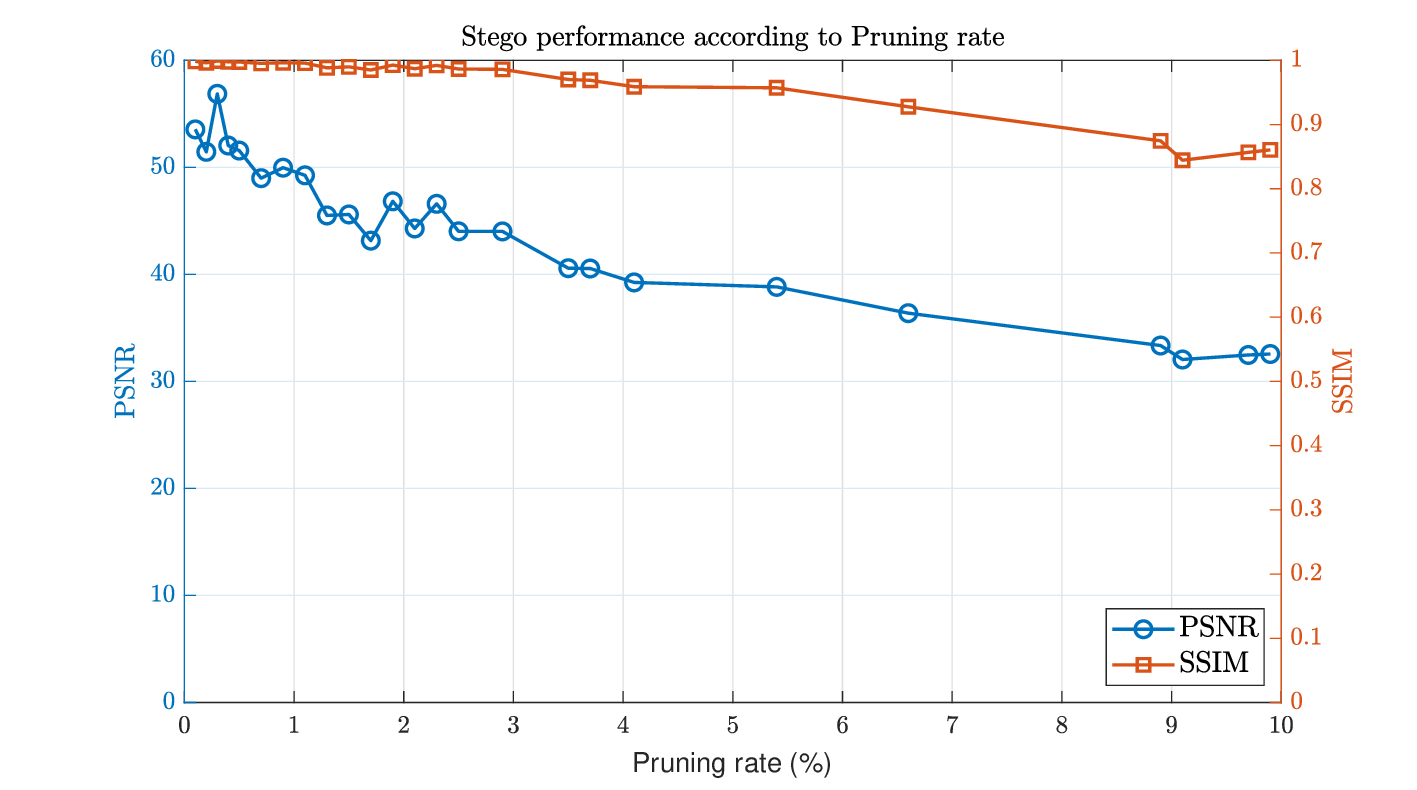}}
\caption{Stego performance according to the pruning rate from 0.1\% to 10\%.}
\label{fig: prune}
\end{center}
\end{figure}

In our manuscript, we assess the robustness of our INRSteg by applying quantization operation. This section introduces another weight compression method, pruning, which is applied to the stego INRs simulating efficient storage of the stego INRs. In order to ensure preserving the quality of reconstruction, we employ iterative magnitude pruning. The weights corresponding to the secret data are masked during pruning, so secret data does not changed during pruning. Pruning is performed on weights with relatively low absolute values among the remaining parts. We initiate the pruning rate at 0.1\%, incrementally increasing it by 0.1\% each time, up to 10\%. Every increasing step, we finetune stego INR with 300 steps with the cover data. \cref{fig: prune} shows the result of stego performance according to the pruning rate. The SSIM values of data reconstructed from stego INR remain high despite a pruning rate of 10\%, indicating successful model compression. Although the PSNR decreases, the values remain sufficiently high, showing that the pruning method effectively compresses the stego INR without significantly compromising the quality of the reconstructed data.
\noindent

\section{Supplementary J}
\label{appendix: 3d limitation}

\begin{figure}[H]
\begin{center}
\centerline{\includegraphics[width=0.7\columnwidth, height=0.4\textwidth]{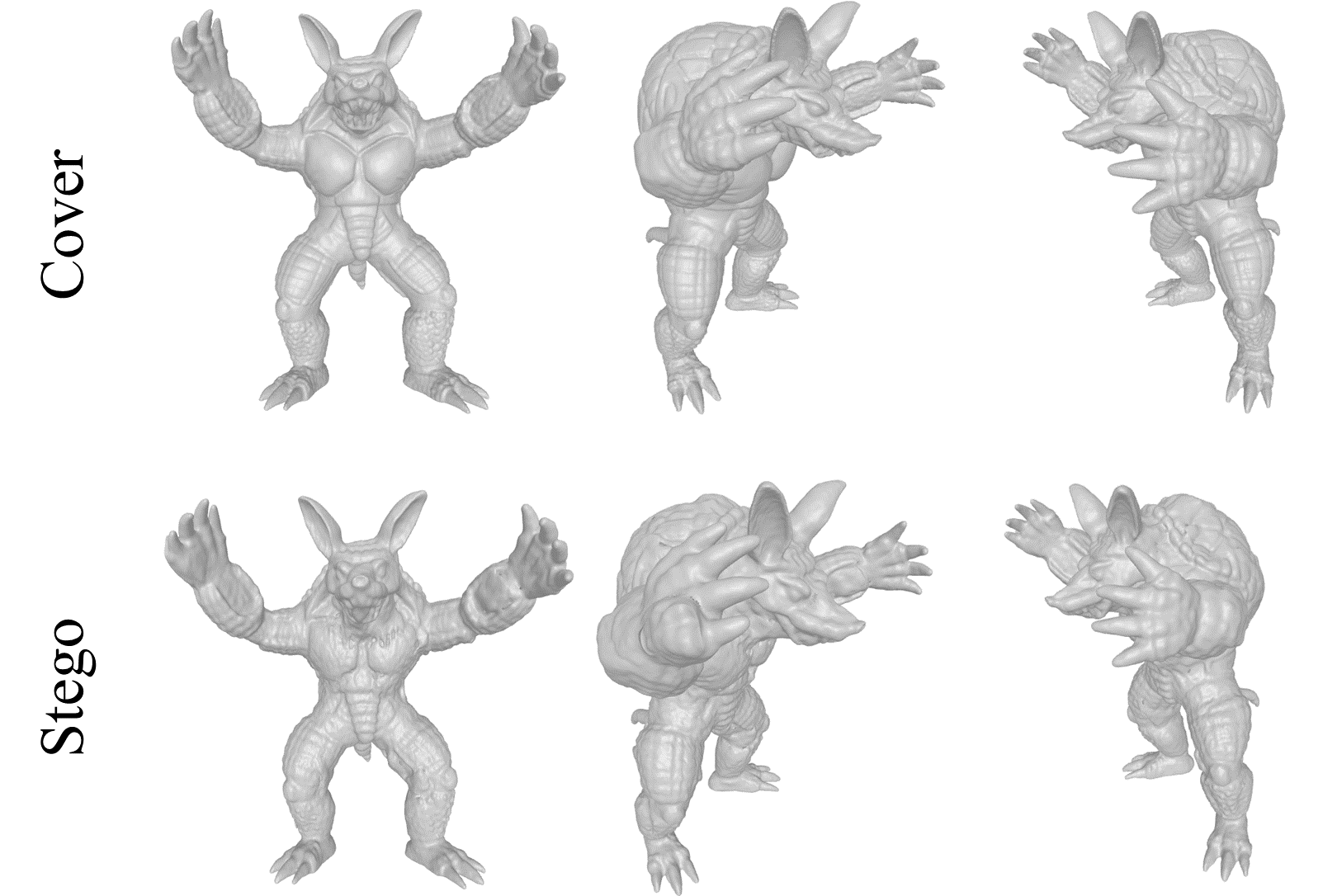}}
\caption{Example of when the stego 3D shape that is not perfectly recovered.}
\label{fig: supple 3d limitation}
\end{center}
\end{figure}

While INRSteg introduces a novel progression in the field of steganography, there is still room for improvement. As SIREN has been retained without any modification as the structure of our INR network, the corresponding limitations also apply to our framework. Due to the underlying unstableness of SIREN, the base structure employed for Implicit Neural Representation (INR), the stego INR may carry defects when the cover modality is 3D shapes as in \cref{fig: supple 3d limitation}. However, this limitation can be resolved by adding additional layers to the stego INR or by employing other advanced structures for INRs, and our framework can be further enhanced by exploring the optimal structure of the neural network and activation function. Another issue is that as INRSteg embraces a a large variety of secret data formats, the metadata plays an important role. Although it hasn't been discussed in the paper, this metadata of secret data can also be easily hidden, disguised as weights of the stego INR in a specific region. 

\noindent

\bibliographystyle{splncs04}